\documentclass[10pt]{iopart}
\usepackage{graphicx}
\usepackage{amssymb}
\usepackage{bbm}
\usepackage[usenames]{color}

\begin{document}

\newcommand{\rep}[2]{\textcolor{red}{\Midline{#1}}\
  \textcolor{OliveGreen}{{#2}}}

\newcommand{\Eq}[1]{Eq.~(\ref{#1})} 
\newcommand{\Fig}[1]{Fig.~\ref{#1}}
\newcommand{\Sec}[1]{Sec.~\ref{#1}}
\newcommand{\EqDef}{\stackrel{\mathrm{def}}{=}}
\newcommand{\UU}{\underline{U}}
\newcommand{\twomat}[4]
   {\left(\begin{array}{cc}#1 & #2 \\ #3 & #4\end{array}\right)}
\def\NP{{\sf{NP}}} 
\def\BPP{{\sf{BPP}}} 
\def\BQP{{\sf{BQP}}}
\def\PBQP{{\sf{PromiseBQP}}}
\def\QMA{{\sf{QMA}}}
\def\TQFT{{\sf{TQFT}}}  
\def\QCD{{\sf{QCD}}}

\newcommand{\Id}{\mathbbm{1}}

\newcommand{\norm}[1]{{\| #1 \|}}  
\newcommand{\abs}[1]{{\left| {#1} \right|}}  
\newcommand{\ket}[1]{{ |{#1} \rangle }}  
\newcommand{\bra}[1]{{ \langle {#1} | }}
\newcommand{\braket}[2]{{ \langle {#1} | {#2} \rangle}}
\newcommand{\ketbra}[2]{{ |{#1} \rangle\langle {#2} | }}

\newcommand{\poly}{\mathrm{poly}}
\newcommand{\TL}{\mathrm{TL}_n(d)}

%
%

\newcommand{\upup}{\ \, \includegraphics[bb=5 5 10 10]{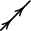}\ \, }
\newcommand{\downdown}{\ \, \includegraphics[bb=5 5 10 10]{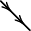}\ \, }
\newcommand{\updown}{\ \, \includegraphics[bb=5 0 10 10]{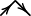}\ \, }
\newcommand{\downup}{\ \, \includegraphics[bb=5 0 10 10]{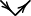}\ \, }
\newcommand{\zigzag}{\ \, \includegraphics[bb=3 0 44 10]{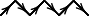}\cdots \, }
\newcommand{\eone}{\ \, \includegraphics[bb=3 5 30 10]{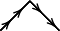} \, }
\newcommand{\ezero}{\ \, \includegraphics[bb=3 0 30 10]{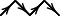} \, }

\newtheorem{thm}{Theorem}[section]
\newtheorem{proposition}{Proposition}[section]
\newtheorem{deff}{Definition}[section] 
\newtheorem{algorithm}{Algorithm}[section] 
\newtheorem{claim}{Claim}[section] 
\newtheorem{lem}{Lemma}[section] 
\newtheorem{conjecture}{Conjecture}[section] 
\newtheorem{hypothesis}{Hypothesis}[section] 
\newtheorem{speculation}{Speculation}[section] 
\newtheorem{notation}{Notation}[section] 
\newtheorem{corol}{Corollary}[section] 
\newtheorem{fact}{Fact}[section] 
\newtheorem{remark}{Remark}[section]

\newenvironment{proof}{\noindent\textit{Proof: }}{$\Box$} 

\title{The $\BQP$-hardness of approximating the Jones Polynomial}
\date{12-December-2010}
\author{Dorit Aharonov and Itai Arad\thanks{\texttt{email:
  itaia@cs.huji.ac.il}}\\
  {\small Department of Computer Science and Engineering,} \\
    {\small Hebrew University, Jerusalem, Israel}}
\maketitle

\begin{abstract}
  A celebrated important result due to Freedman, Larsen and Wang
  \cite{ref:Fre02} states that providing additive approximations of
  the Jones polynomial at the $k$'th root of unity, for constant
  $k=5$ and $k\ge 7$, is $\BQP$-hard. Together with the algorithmic
  results of \cite{ref:Fre02b, ref:Aha05}, this gives perhaps the most
  natural $\BQP$-complete problem known today and motivates further
  study of the topic.  In this paper we focus on the universality
  proof; we extend the result of \cite{ref:Fre02} to $k$'s that
  grow polynomially with the number of strands and crossings in the
  link, thus extending the $\BQP$-hardness of Jones polynomial
  approximations to all values for which the AJL algorithm applies
  \cite{ref:Aha05}, proving that for all those values, the problems
  are $\BQP$-complete. As a side benefit, we derive a fairly
  elementary proof of the Freedman \emph{et al.} density result
  \cite{ref:Fre02}, without referring to advanced results from Lie
  algebra representation theory, making this important result
  accessible to computer science audience.  We make use of two
  general lemmas we prove, the \emph{bridge lemma} and the
  \emph{decoupling lemma}, which provide tools for establishing
  density of subgroups in $SU(n)$.  Those tools seem to be of
  independent interest in more general contexts of proving quantum
  universality.  Our result also implies a completely classical
  statement, that the \emph{multiplicative} approximations of the
  Jones polynomial, at exactly the same values, are
  $\#\mathsf{P}$-hard, via a recent result due to Kuperberg
  \cite{ref:Gre09}.  Since the first publication of those results in
  their preliminary form \cite{ref:Aha06}, the methods we present 
  here were used in several other contexts \cite{ref:Aha07,
  ref:Sho08}. This paper is an improved and extended version of the
  results presented in \cite{ref:Aha06}, and also includes
  discussions of the developments since then. 
\end{abstract}

\section{Introduction} 

What is the computational power of quantum computers?  This question
is fundamental both from a computer scientist as well as a physicist
points of view. This paper attempts to improve our understanding of
this question, by studying perhaps the most natural $\BQP$-complete
problem known to us today: the problem of approximating the Jones
polynomial. Here we try to clarify the reasons for its
$\BQP$-hardness, as well as extend its applicability, and on the
way, gain better understanding and new tools for proving quantum
universality in general.

\subsection{Background} 

The Jones polynomial, discovered in $1985$ \cite{ref:Jon85}, is a
very important knot invariant in topology; it assigns a one variable
Laurent polynomial $V_L(t)$ to a link $L$, in such a way that
isotopic links are assigned the same polynomial. It is an extremely
difficult object to compute -- evaluating it at any point except for
a few trivial ones is \textsf{\#P}-hard \cite{ref:Jae90}.  The
importance of the Jones polynomial was manifested in connections to
numerous areas in mathematics, from the statistical physics model
known as the Potts model to the study of DNA folding. Among its many
connections, an extremely important one was drawn by Witten in 1989
to quantum mechanics, and, specifically, to Topological Quantum
Field Theory ($\TQFT$) \cite{ref:Wit89}. Witten showed how the Jones
Polynomial naturally appears in the Wilson lines of the $SU(2)$
Chern-Simons Topological Quantum Field Theory.

About a decade later, $\TQFT$ entered the scene of quantum
computation when Freedman suggested a computational model based on
this theory \cite{ref:Fre98}. The works of Freedman, Kitaev, Larsen 
and Wang \cite{ref:Fre02, ref:Fre02b, ref:Fre03} showed an 
equivalence between the $\TQFT$ model and the standard model of
quantum computation. On one hand, they gave an efficient simulation
of $\TQFT$ by a quantum computer \cite{ref:Fre02b}.  On the other
hand, they showed that quantum computation can simulate $\TQFT$
efficiently \cite{ref:Fre02}.  These results draw interesting
connections between quantum computation and the Jones polynomial.
The simulation of $\TQFT$ by quantum computers implicitly implies
(via the results of Witten) the existence of a quantum algorithm for
{\it approximating} the Jones polynomial evaluated at the fifth root
of unity $t=\exp(2\pi i/5)$, to within a certain \emph{additive}
approximation window. In the other direction, the simulation of
quantum computers by $\TQFT$ implicitly implies that the same Jones
polynomial approximation problem (with the same additive
approximation window) is $\BQP$-hard; the proof uses Lie algebras
extensively. This draws an important equivalence between the two
seemingly completely different problems of quantum computation and
the approximation of the Jones polynomial of links.  

The above mentioned important results were stated in the language of
$\TQFT$, and relied on advanced results from Lie algebras theory;
this made the results inaccessible for much of the computer
science community for a while.  In \cite{ref:Bor05}, clear
statements of the results were provided using a computational
language, but without proofs; an explicit algorithm was thus still
missing, as well as a proof from first principles of universality.  

Few years ago, Aharonov, Jones and Landau \cite{ref:Aha05} provided
an explicit and efficient quantum algorithm for the problem of
approximating the Jones polynomial of a given link, at roots of
unity of the form $\exp(2\pi i/k)$, using the standard quantum
circuit model. The algorithm uses a combination of simple to state
combinatorial and algebraic results of over $20$ years ago due to
Jones. The main ingredient is a certain matrix representation,
called the \emph{path-model representation}, which maps elements
from an algebra of braid-like objects (called the 
\emph{Temperley-Lieb Algebra} $\TL$), to operators acting on
\emph{paths} of $n$ steps on a certain graph $G_k$.  In the cases in
which this representation is unitary, this gives a simple-to-state
quantum algorithm for the approximation of the Jones polynomial: the
matrices are applied by the quantum computer, and the approximation
of the Jones polynomial is derived by approximating a certain trace
of the resulting unitary matrix. This bypasses the $\TQFT$ language
altogether.  

The universality proof due to \cite{ref:Fre02}, stated first in 
terms that were also closer to the $\TQFT$ language, can also be
made explicit in the standard quantum model language, without
referring to $\TQFT$. This can be done using a mapping suggested by
Kitaev \cite{ref:Kit05}, and independently Wocjan and Yard
\cite{ref:Woc06a}, in which the basis states of one qubit are
encoded by one of two possible paths of length four in the space of
the path model representation. 

The results described above imply an explicit proof in the standard
quantum computation model that the problem of approximating the
Jones polynomial at the fifth root of unity, and in fact, for any
primitive root of unity $\exp(2\pi i/k)$, for constant $k>4, k\ne 6$
is $\BQP$-complete and thus equivalent in a well-defined sense to
standard quantum computation.  This is arguably the most natural
$\BQP$-complete problem known to us today, (though see
\cite{ref:Woc06b, ref:Woc06c, ref:Woc07}).  The fact that the
problem is $\BQP$-complete, highlights the importance of this
problem in the context of quantum computational complexity, and
motivates deeper investigation of the intriguing connections and
insights revealed by those results.  

We remark that, as is usually done in the literature, we
slightly abuse notation and when we say a problem is 
$\BQP$-complete, we in fact mean this in the context of promise 
problems; just like in the case of $\BPP$, there are no known
$\BQP$-complete problems in the strict sense of the term, and so we
actually mean that the problem is $\PBQP$-complete. For a detailed
discussion of this point, see \cite{ref:PromiseBQP,ref:PromiseBPP}
and references therein. 

One natural direction to pursue is to try and generalize the
algorithm in various directions. Several results extended the Jones
polynomial approximation algorithms to other knot invariants and to
more general braid closures (see, e.g., \cite{ref:Sil06,
ref:Woc06a}) to evaluation of the Potts model partition function and
the Tutte polynomial \cite{ref:Aha07}, and to approximations of
tensor networks \cite{ref:Ara10}, as well as to the Turaev-Viro
invariant \cite{ref:Gor10}. In this paper we take the other
direction: we attempt to study and further clarify the reasons for
the $\BQP$-hardness of those problems, and expand its range of
applicability, with the hope of clarifying the source of the
computational power of quantum computation.

\subsection{Results and Implications} 

We ask here the following natural question.  It turns out that the
algorithms given in the work of Aharonov et al. work not only for
constant $k$, but also for asymptotically growing $k$'s.  To be more
precise, \cite{ref:Aha05} gives an efficient quantum algorithm to
approximate the Jones polynomial of a certain closure (called the
plat closure) of an $n$-strands braid with $m$ crossings, evaluated
at a primitive root of unity $\exp(2\pi i/k)$.  The running time of
the algorithm is polynomial in $m,n$ and $k$.  The algorithm is
therefore efficient even if $k$ grows polynomially with $n$.  On the
other hand, the proof of $\BQP$-hardness is only known to hold for
constant $k$.  Therefore, in \cite{ref:Aha05} the following natural
question was raised: what is the complexity of approximating the
Jones polynomial for polynomially bounded $k$?  It was left open
whether it is $\BQP$-hard, doable in $\BPP$, or maybe somewhere in
between. 

In this paper we resolve this question, and show that for any
polynomially bounded $k$, the problem is $\BQP$-hard. The following
is a rough statement of the result; exact statement is given in
Theorem~\ref{thm:poly-k} in \Sec{sec:poly-k}.
\begin{thm}
\label{thm:intro} 
  The problem of approximating the Jones polynomial of the plat
  closure of a given braid $b$, with $m$ crossings, at $\exp(2\pi
  i/k)$, where both $m$ and $k$ are polynomially bounded in $n$, to
  within the same accuracy as is done in \cite{ref:Aha05}, is
  $\BQP$-complete.
\end{thm} 

We thus show that in all cases where the AJL algorithm
\cite{ref:Aha05} is known to be efficient, we derive that the
problem it solves is $\BQP$-complete.  The proof is not a mere
extension of the previous constant $k$ case, and there are severe
problems to overcome. 

As a side benefit, our proof also simplifies the original proofs for
the constant $k$ case \cite{ref:Fre02}, and reproves it almost from
first principles, without using advanced results from Lie algebra,
thus making it more accessible to the computer science audience.
Indeed, since the preliminary publication of the results presented
here in \cite{ref:Aha06}, the methods we developed here were applied
in several other contexts (See \Sec{sec:related}).

We will soon outline the general approach towards the proof of
universality of the constant $k$ case, the difficulties in extending
the proof to non-constant $k$, and our methods to overcome them. 
Before that, let us mention interesting connections and further
implications to the complexity of \emph{multiplicative}
approximations of the Jones polynomial. 

\subsection{Implication to hardness of the multiplicative approximation 
problem} 

A significant ``drawback'' of the AJL algorithm is the fact that it
provides an \emph{additive approximation} to the Jones polynomial.
It can approximate the Jones polynomial up to an additive error of
$\Delta/\poly(n)$, with $\Delta$ being some scale (which is easy to
calculate). The problem is that the exact value of the polynomial
might be exponentially smaller than $\Delta$, making this kind of
additive approximation useless. A partial answer to this
``drawback'' is found in its complementing result, the
$\BQP$-hardness theorem, which we re-prove in this paper; it shows
that despite the seeming weakness of the approximation, it is as
hard as the hardest problems that a quantum computer can solve.
Thus, there exist links for which the additive approximation of the
AJL is non-trivial. Nevertheless, one can rightfully argue that the
situation is still not satisfactory; additive approximations are far
less interesting from an algorithmic point of view, and we would
have liked to focus on a much better and more natural approximation
notion, namely a multiplicative one. 

Goldberg and Jerrum \cite{ref:GJ} studied the complexity of 
multiplicative approximations of the Tutte polynomial.  The Jones
polynomial (of {\it alternating} links) is a special case of this
important polynomial.  Their results imply that the multiplicative
approximation (to within a constant arbitrarily close to $1$) of the
Jones polynomial of alternating links, at certain real values, is
$\NP$ hard (relative to $\mathsf{RP}$)\footnote{They have also shown
that in some special cases, multiplicative approximations of the
Tutte polynomial are $\#\mathsf{P}$-hard, but these cases do not
correspond to the Jones polynomial.} Those values, however, do not
intersect the values for which $\BQP$ additive approximations exist
due to the AJL algorithm, as they only apply to real points, while
the AJL works at the complex roots of unity.  And so one might still
hope that the AJL algorithm as well as the universality proofs for
those values can be improved and stated using the multiplicative
approximation notion.  

A beautiful recent result by Kuperberg \cite{ref:Gre09}, helps to
shed light on this matter. Kuperberg observed that $\BQP$ hardness
of additive approximations seems to go {\it hand in hand} with the
$\#\mathsf{P}$-hardness of the \emph{multiplicative approximation}
at the same values, via the result of Aaronson that
$\mathsf{PostBQP}=\mathsf{PP}$ \cite{ref:Aar05}, as well as on the
exponential efficiency of the Solovay-Kitaev algorithm. Using these
ideas, Kuperberg proved that the multiplicative approximations of
the Jones polynomial of a plat closure of a braid, evaluated at the
$k$th root of unity, for constant $k=5$ and $k\ge 7$, are
$\#\mathsf{P}$-hard.  Essentially, the argument is that by
Aaronson's result, in order to solve $\#\mathsf{P}$-hard problems,
it suffices to be able to compute, or even provide multiplicative
approximations of, conditional probabilities for the outputs of a
given quantum circuit. However, for $k$ for which $\BQP$-hardness 
of additive approximations of the Jones polynomial holds, we can use
the same mapping from circuits to links used for the 
$\BQP$-hardness to derive {\it exponentially} good approximations of
those conditional probabilities in terms of the Jones polynomial of
some link, where the link need only be polynomial in the number of
gates in the circuit, due to the exponential efficiency of the 
Solovay-Kitaev theorem. Note that the final result is a purely
classical result that is derived using quantum complexity tools.  It
turns out that the argument goes through also for the universality
proofs in this paper, and hence, we get the following corollary: 

\begin{corol}    
  The problems of the approximation of the Jones polynomial at the
  same points and parameters for which theorem~\ref{thm:intro}
  implies $\BQP$-hardness, with the approximation replaced
  by \emph{multiplicative approximation} to within a constant
  arbitrarily close to $1$, are 
  $\#\mathsf{P}$-hard. 
\end{corol} 

We now proceed to outline the proof of Theorem \ref{thm:intro}. 
Let us start with explaining the constant $k$ case first.  

\subsection{Proof Outline of the Constant $k$ case} 

\begin{figure}
  \center
  \includegraphics[scale=1]{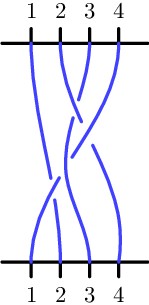}
  \caption{An example of a 4-strands braid}
 \label{fig:4strand}
\end{figure}

Given an algorithm that calculates the Jones Polynomial of any link
at $\exp(-2\pi i/k)$ (for some integer $k>4$ and $k\ne 6$) in
polynomial time in the number of crossings in the link, and a
classical Turing machine - we can simulate a Quantum computer
efficiently. How is that possible? The key idea, which is used also
in the algorithmic result of \cite{ref:Aha05}, the existence of an
intimate connection between two, seemingly distinct, worlds: links
and unitary matrices. The connection is the so-called ``path-model
representation'', which is defined for every integer $k$.  The $k$th
path model representation maps every $n$-strand braid $b$ in the
braid group (e.g., \Fig{fig:4strand}) into a unitary matrix
$\rho(b)$.

\begin{figure}
  \center
  \includegraphics[scale=1]{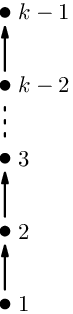}
  \caption{The graph $G_k$}
 \label{fig:Gk}
\end{figure}

$\rho(b)$ acts on a Hilbert space spanned by \emph{paths} of length
$n$ on a certain graph, $G_k$, which is simply the line graph of
$k-2$ segments (see \Fig{fig:Gk}). 

\begin{figure}
  \center
  \includegraphics[scale=1]{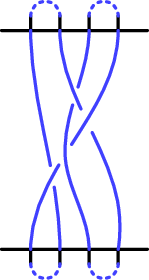}
  \caption{The plat closure of the 4-strand braid from
  \Fig{fig:4strand}}
 \label{fig:plat}
\end{figure}

As was shown by Jones \cite{ref:Jon85, ref:Jon86}, the unitary matrix
$\rho(b)$ can be related to the Jones polynomial of the link
$b^{pl}$ derived from the braid $b$ by closing its strands in a
certain way called the \emph{plat closure} (see example in Figure
\ref{fig:plat});

The connection is that the expectation value
$\bra{\alpha}\rho(b)\ket{\alpha}$ (where $\ket{\alpha}$ is some
special state) is proportional to the Jones polynomial $V_{b^{pl}}$
of the plat closure of the braid $b$, evaluated at the $k$th root of
unity (with an easy to calculate proportionality constant). To prove
$\BQP$ hardness of the approximation of the Jones polynomial, it
thus suffices to prove $\BQP$ hardness of the approximation of
$\bra{\alpha}\rho(b)\ket{\alpha}$ for a given braid $b$.  

The strategy to do this is to show that any given quantum circuit,
namely a sequence of gates $U_L\cdot U_{L-1}\cdots U_1$, can be
mapped to a braid $b$, such that the value $\bra{0^n}U_L\cdot
U_{L-1}\cdots U_1\ket{0^n}$ is proportional to
$\bra{\alpha}\rho(b)\ket{\alpha}$. 

The $\BQP$-hardness proof thus boils down to showing that a general
quantum gate can be approximated efficiently using the unitary
images of braids by the path-model representation. More precisely,
one considers some subset of the generators of the braid group (each
generator is simply a crossing of two adjacent strands). Each such
generator is mapped to a certain ($k-$dependent) unitary operators
on the space of paths. The main difficulty in the proof is to show
that the group generated by the images of those generators is dense
in a large enough subgroup of the unitary group, to contain all
unitary gates. Once this is shown, it is standard to apply the
famous Solovay-Kitaev theorem \cite{ref:Kit02a} to show that
\emph{density implies efficiency}. In other words, once the subgroup
is dense, then Solovay-Kitaev gives a method to approximate every
gate in the quantum circuit by a polynomially bounded in length
sequence of generators, and universality follows.  

But how does one prove the density? The starting point is 
Kitaev-Wocjan-Yard four steps encoding \cite{ref:Kit05, ref:Woc06a}
which encodes the state of one qubit into four steps paths. For two
qubits, these paths correspond to $8$-strands braids, $4$ for each
encoded qubit. In fact, the four dimensional Hilbert space of the
two qubits is encoded into a space spanned by $4$ paths, which is 
embedded into an ``invariant'' space spanned by all $14$ paths on
the $8$ strands; see \Sec{sec:analyze} and \Fig{fig:8p} for the
details.  Density thus means that we can approximate any matrix in
$SU(14)$ (and thus also any matrix in $SU(4)$ embedded in it), using
our $k$-dependent generators.  In order to prove density, the idea
is to first restrict attention to some two dimensional subspace, and
show density in $SU(2)$. This was essentially done by Jones
\cite{ref:Jon83}.  We then gradually increase the dimensionality of
the space on which we have density, to $SU(14)$, by adding one or
more dimensions at a time; to this end we introduce two lemmas which
are useful tools for proving universality in general: the
\emph{Bridge Lemma} \ref{thm:mix} and the \emph{Decoupling Lemma}
\ref{thm:decouple}. We explain those later in the introduction, in
\Sec{sec:lemmas} since they are of independent interest.  Using
these lemmas, we can build up our way from density on $SU(2)$ to the
desired density on $SU(14)$; this completes the density proof of the
constant $k$ case. We get an almost self-contained, fairly
elementary proof.

\subsection{Proof outline of the polynomially growing $k$ case}

We would now like to move to the asymptotically growing $k$ case. 
Here, however, there is a subtle point in the above line of
arguments. Indeed, density still holds.  But the step of
\emph{density implies efficiency} fails. The starting point of the
Solovay-Kitaev theorem is the construction, using the set of
generators, of an $\epsilon$-net in the unitary group, where
$\epsilon$ is some small enough constant. Such an$\epsilon$-net is
easy to construct, given a \emph{fixed} set of generators that span
a dense subgroup - essentially, brute force would do the trick.
More precisely, one considers an arbitrary delta-net in the unitary
group $SU(14)$, for delta being $\epsilon$/3; such a net contains a
finite number of points. Due to the density, by brute force we can
find delta approximations of all those finitely many points by
finite products of our generators, and those products constitute the
$\epsilon$-net. The complexity of this initial step might be
horrible, but it depends only on $k$ and $\epsilon$, and not on $n$;
for a fixed $k$, it is thus constant. However, if $k$ is
asymptotically growing in $n$, then so are the generators. The brute
force procedure might depend in an uncontrollable way on $k$, and
thus on $n$.  It is therefore no longer clear that the very first
step of the Solovay-Kitaev theorem, that of creating the
epsilon-net, can be done efficiently. 

We give here a very rough sketch of how we overcome this difficulty.
Looking at the $k$-dependence of the generators, we see that as
$k\to \infty$, their eigenvectors converge to a fixed limit, while
their corresponding eigenvalues behaves as $\exp(-2i\pi/k)$. The
idea then is to fix a $k_0$ and to consider special auxiliary
generators: generators whose eigenvectors coincide with the $k\to
\infty$ limit, but their eigenvalues are the fixed $k_0$
eigenvalues, $\exp(-2i\pi/k_0)$. This set of auxiliary generators is
independent of $n$, and we show it too spans a dense subgroup in
$SU(14)$; thus we can construct an $\epsilon$-net from it using a
straightforward brute-force search. For every sufficiently large
$k$, the eigenvectors of the $k$-dependent generators would be close
enough to those of the limit $k\to\infty$, and thus to the
eigenvectors of the auxiliary generators; by taking the $k/k_0$'s
power of the of the $k$ dependent generators, we get the $k/k_0$'s
power of their eigenvalues $\exp(-2i\pi/k)$ and thus we approximate
the $\exp(-2i\pi/k_0)$ eigenvalues of the auxiliary generators.  For
large enough $k$ the eigenvectors of $k$ would be close enough to 
the $k\to\infty$ eigenvectors, and the truncation error when
approximating $k/k_0$ by an integer would be negligible, and so we
get an approximation of the auxiliary generators by $k$-dependent
generators.  We can now substitute these approximations in the
$\epsilon$-net made of the auxiliary generators, to get an efficient
construction of an $\epsilon$-net consisting of the $k$-generators.
We can now apply the Solovay-Kitaev theorem using this net. 

\subsection{Tools for Universality: The Bridge lemma and the Decoupling lemma}
\label{sec:lemmas}

We provide here the rough statements of the two lemmas we use here
for proving density, since they seem to be useful for proving
universality in a variety of other contexts.  

The bridge lemma roughly says that if we have density in the unitary
groups acting on two orthogonal subspaces, $A$ and $B$, with
$dim(B)>dim(A)$ and an additional unitary which mixes the two
subspaces (in some well defined sense), we also have density on the
direct sum of the spaces. This general lemma is very reminiscent of
a lemma which appeared in an early version of
\cite{ref:Aha97,ref:FT}.  Its proof is based on simple linear
algebra, and is iterative; it uses a combination of ideas by
Aharonov and Ben-Or \cite{ref:Aha97,ref:FT} and by Kitaev
\cite{ref:Kit02a}.

The decoupling lemma deals with the following scenario: a certain
subgroup of the unitary matrices can be shown to be dense when
restricted to one subspace and also to another subspace orthogonal
to it.  When we want to combine the two spaces, we encounter a
problem since there may be correlations between how the matrices act
on the two subspaces. The lemma states that if the dimensions of the
spaces are different, it is possible to ``decouple'' those
correlations and approach any unitary on one space while 
approaching the identity on another, and vice verse.  The proof of
the decoupling lemma uses simple analysis.

\subsection{Related work and discussion}
\label{sec:related}

Since the first publication of the results presented here (in
preliminary form) \cite{ref:Aha06}, they were already used in
several contexts: Shor and Jordan \cite{ref:Sho08} built on the
methods we develop here to prove universality of a variant of the
Jones polynomial approximation problem, in the model of quantum
computation with one clean qubit. In the extension of the AJL
algorithm \cite{ref:Aha05} to the Potts model \cite{ref:Aha07},
Aharonov \emph{et al.} build on those methods to prove universality
of approximating the Jones polynomial in many other values, and even
in values which correspond to non-unitary representations.  We hope
that the method we present here will be useful is future other
contexts as well. 

Finally, we mention that the results of this paper should be viewed
in a somewhat wider context of the notion of quantum ``encoded
universality''.  By that we mean the following: rather than showing
that a set of gates on $n$ qubits generates a dense subgroup in the
unitary group on those $n$ qubits, as is done in the standard notion
of quantum universality, one proves that the set of gates in fact
generates a dense set in the unitary group on a space of dimension
less than $2^n$, which is \emph{embedded} or \emph{encoded} in the
bigger $2^n$ dimensional Hilbert space.  If the encoding can be
computed efficiently, and the encoded Hilbert space is of large
enough dimension, this suffices for efficient simulation of 
universal quantum computation.

In fact, though not explicitly stated, encoded universality is
exactly what was proved by Freedman \emph{et al.} in their original
universality proof of the $\TQFT$ simulation \cite{ref:Fre02}, and
of course in the universality proofs based on them \cite{ref:Kit05,
ref:Woc06a} including this current paper. The first time encoded
universality was used \cite{ref:Fre02} can probably be tracked to
the proof that real quantum computation suffices to simulate all of
quantum computation by Bernstein and Vazirani \cite{ref:bv}.  This
notion was also used in various other contexts, e.g., in the context
of fault tolerance and decoherence free subspaces \cite{ref:Bac00}
as well as in the encoded universality proof of the Heisenberg
interaction \cite{ref:Bac01, ref:Kem01}.  In this paper we in fact
provide general tools to prove density for such encoded universality
scenarios. 

{~}

\noindent\textbf{Organization of the paper:}\\ 
In \Sec{sec:background} we provide the required mathematical
background on links, braids, Temperley-Lieb algebra and the
path-model representation that is needed for the proof. In
\Sec{sec:const-k} we state and prove the constant-$k$ universality
theorem by using the density and efficiency theorem. This theorem,
which is the heart of the proof, is proved separately in
\Sec{sec:B8-const}. In \Sec{sec:poly-k} we state and prove the main
result of this paper, the $\BQP$-hardness of the $k=\poly(n)$ case.
Finally, in \Sec{sec:tools} we prove the bridge and decoupling
lemmas that are used in the density proof in \Sec{sec:B8-const}.

\section{Background: braid groups, the Temperley-Lieb algebra and
  path-model representations}
\label{sec:background}

In this section we give a brief overview of the algebraic and
topological definitions and tools that we need to prove
theorem~\ref{thm:intro}. We define the braid group, its embedding in
the Temperley-Lieb algebra, and the path-model representation and
its relation to the Jones polynomial. A more detailed description of
these subjects based on first principles can be found in
\cite{ref:Aha05, ref:Aha07}.

\subsection{The braid group $B_n$}

Loosely speaking, a braid is a set of $n$ strands that connect two
horizontal bars, such that each strand is tied exactly to one peg on
the top bar and one peg on the bottom bar. When drawing the braid
schematically on a paper, the strands may pass over and under each
other, but at any point they must not be completely horizontal.
Braids which can be deformed into each other without tearing any of
the strands are considered identical. An illustration of a 4-strand
braid is given in \Fig{fig:4strand}.

The set of all braids with $n$ strands forms an infinite and discrete
group which is called the \emph{braid group $B_n$}. The product rule
for $b_1 b_2$ is defined by placing the braid $b_1$ above the braid
$b_2$ and fusing the bottom of the $b_1$ strands with the top of the
$b_2$ strands. The identity element is the braid with $n$ straight
lines that connect each peg at the bottom bar to its corresponding
peg at the upper bar. 

\begin{figure}
  \center
  \includegraphics[scale=1]{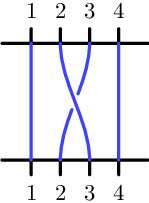}
  \caption{The generator $\sigma_2$ in the braid group $B_4$.}
 \label{fig:sigma-i}
\end{figure}

In 1925, Artin proved that $B_n$ admits a finite presentation (the
\emph{Artin presentation}) \cite{ref:Art25}, with $n-1$ generators
$\{\sigma_i\}$ that satisfy the following constraints:
\begin{eqnarray}
  \label{eq:def-B1}  
  \sigma_i\sigma_j &= \sigma_j\sigma_i \quad 
    \mbox{for $|i-j| \ge 2$} \ , \\
  \sigma_i\sigma_{i+1}\sigma_i &= \sigma_{i+1}\sigma_i\sigma_{i+1} \ .
  \label{eq:def-B2}  
\end{eqnarray}
Pictorially, $\sigma_i$ is a braid that is identical to the unity
braid in all strands except for the $i$ and $i+1$ strands which
cross each other once (the $i+1 \to i$ strand goes over the $i\to
i+1$ strand), connecting the lower $i$'th peg to the upper $i+1$ peg
and vice verse. The diagram of $\sigma_2$ in $B_4$ is given in
\Fig{fig:sigma-i}. It is an easy exercise to verify graphically that
the braid generators indeed satisfy (\ref{eq:def-B1}),
(\ref{eq:def-B2}). 

\subsection{From braids to links}
\label{sec:plat}

A link is an embedding of one or more closed loop in
$\mathbbm{R}^3$. We first notice that a braid can be transformed
into a link by connecting its open endpoints. Such an operation is
called a \emph{closure}, and here we focus on one particular
closure: the \emph{plat closure}. This closure is defined only for
braids with an even number of strands. It is the link that is formed
by connecting the top pegs with odd numbers with the peg to their
right, and doing the same with the bottom pegs. The plat closure of
a braid $b\in B_n$ is denoted by $b^{pl}$. Figure~\ref{fig:plat}
shows the plat closure of the 4-strand braid from \Fig{fig:4strand}.

\subsection{The Temperley-Lieb Algebra $\TL$}
\label{sec:TL}

We are interested in defining certain useful representations for the
braid group $B_n$, which we will later relate to the Jones
polynomial. To this end, we first consider the Tempreley-Lieb
Algebra $\TL$ \cite{ref:Tem71}. This is because the generators
$\sigma_i$ of the braid group $B_n$, and therefore all of $B_n$, can
be embedded in that algebra. Hence, any representation of $\TL$
yields a representation of $B_n$.

\begin{figure}
  \center
  \includegraphics[scale=1]{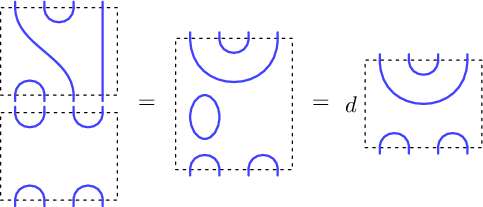}
  \caption{Multiplying tangles in the $\TL$ algebra. The first
  diagram is put on top of the second, and the pegs are connected.
  In the resulting tangle, every loop is removed and replaced with an
  over all $d$ factor.}
 \label{fig:tangles}
\end{figure}

For any scalar $d$, the $\TL$ algebra is an algebra of tangle
diagrams that, much like braid diagrams, connect $n$ lower pegs to
$n$ upper pegs. However, unlike the case of braid diagrams, here we
do not allow crossings, but we do allow horizontal lines, including
local minimas and maximas. Finally, closed loops are not allowed. 
To multiply two tangles, we put one on top of the other, connecting
lower pegs with upper pegs. Any closed loop that is created in this
process is then taken out of the diagram and replaced with an
overall factor of $d$, called the \emph{loop value}.  See
\Fig{fig:tangles} for an example.

\begin{figure}
  \center
  \includegraphics[scale=1]{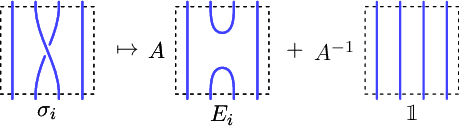}
  \caption{The embedding of the braid group $B_n$ in the
  Temperley-Lieb algebra $\TL$. The generator $\rho_i$ is mapped
  into a superposition of the tangles $E_i$ and $\Id$, with $A$
  given by $-(A^2+A^{-2})=d$.}
 \label{fig:cross-opening}
\end{figure}

The braid-group
$B_n$ can be embedded in the $\TL$ algebra using the following map,
shown schematically in \Fig{fig:cross-opening}:
\begin{equation}
\label{def:TL-sigma}
  \sigma_i \to AE_i + A^{-1}\Id \ .
\end{equation}
Here, $\Id$ is the identity tangle -- the tangle that connects every
lower $i$'th peg to the corresponding upper $i$'th peg. $E_i$ is the
tangle that is form by a ``cap'' that connects the lower $i$,$i+1$
pegs and a ``cup'' that connects the corresponding upper pegs, and
the reset of the pegs are connected by identity lines. Finally, $A$
is the scalar defined by 

\begin{equation}\label{eq:dA}
d=-(A^2+A^{-2}).
\end{equation} 
It is an easy exercise
to verify that the $\sigma_i$ defined by \eref{def:TL-sigma} indeed
satisfy the Artin presentation (\ref{eq:def-B1},~\ref{eq:def-B2}).
 
It follows that any matrix representation of the $\TL$ algebra
yields a matrix representation of the braid group $B_n$.
We will next construct the representations which we will be using.

\subsection{The path-model representation}
\label{sec:path-rep}

The path-model representations are a family of representations for
the Temperley-Lieb algebras \cite{ref:Tem71} that induce
representations for the braid group $B_n$ via \eref{def:TL-sigma}.
They were constructed in \cite{ref:Jon85,ref:Jon86}, and form the
basis of the AJL algorithm \cite{ref:Aha05}. Here we will provide
just minimal details that are needed to understand the use of these
representations when applied to the braid group. A broader
presentation of this beautiful subject, together with its relation
to the Temperley-Lieb algebras and the knot invariants, can be found
in \cite{ref:Aha05, ref:Aha07}. 

We work with sub-family of the path-model representations, which is
characterized by an integer $k\ge 3$ and yields a representation for
$\TL$ with 
\begin{equation}\label{def:d0}
  d=2\cos(\pi/k) \ . 
\end{equation}
When applied to the braid group via \eref{def:TL-sigma}, using 
\begin{equation}
\label{def:A0}
  A=ie^{-i\pi/(2k)} \ ,
\end{equation}
(which satisfies \eref{eq:dA}) this representation becomes a
\emph{unitary} representation of $B_n$. The image of every tangle
$T\in \TL$ (or $b\in B_n$) under this representation is denoted by
$\rho(T)$ (or $\rho(b)$) and it acts on a finite Hilbert space. To
understand the structure of this space, we introduce the graph
$G_k$, which is made from a set of $k-1$ sites (vertices) and $k-2$
edges that connect them. The sites are ordered from bottom to top
one above the other, as described in \Fig{fig:Gk}. To each site we
assign a number according to its position, starting with $1$ at the
bottom.

We then consider all possible $n$-steps walks (paths) over the graph
$G_k$ that start at site $1$ \emph{and never leave $G_k$}. We use
these paths to define the Hilbert space $H_{n,k}$ of $n$-steps paths
over $G_k$: every path $p$ is mapped to a vector $\ket{p}\in
H_{n,k}$, and we define the set of all paths to be an orthonormal
basis of $H_{n,k}$.
\begin{figure}
  \center \includegraphics[scale=1]{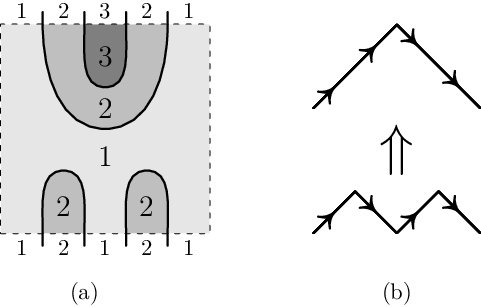} 
  
  \caption{An example of a tangle and two compatible paths. Here the
    lower path $p=1\to 2\to 1\to 2\to 1$ is shown to be compatible
    with the upper path $p'=1\to 2\to 3\to 2\to 1$. These paths
    define a unique labeling of every region in the tangle by a
    vertex of $G_k$. } 
    \label{fig:coloring}
    \end{figure}

To define $\rho$, we will describe the action of $\rho(T)$ on some
$\ket{p}\in H_{n,k}$, where $T\in \TL$. $\rho(T)\ket{p}$
is a linear combination of paths. A path $p'$ with a non-vanishing
weight in that combination is said to be \emph{compatible} with $p$
with respect to $T$. To decide whether $p'$ is compatible with $p$,
we first draw $T$ in a box. The $n$ lower pegs divide the lower
boundary of the box into $n+1$ segments, which we call lower gaps,
and similarly the upper pegs define $n+1$ upper gaps. We now
associate every vertex of the path $p$ with the lower gaps (starting
from the left-most gap, which must be 1), and the upper gaps with
$p'$. We notice that as $T$ contains no loops, it partitions the
box into non-overlapping regions, and each region must be connected
to at least one gap (either lower or upper). Therefore every region
in the box is associated with at least one vertex, either of $p$ or
of $p'$ (or of both).  Then $p$ and $p'$ are compatible iff every
region is associated with exactly one vertex. When this happens, the
paths define a ``labeling'', or a ``coloring'' of the regions. An
example of two compatible paths and the coloring they defined is
shown in \Fig{fig:coloring}. There, the path $p=1\to 2\to 1\to 2\to
1$ is shown to be compatible with the path $p'=1\to 2\to 3\to 2\to
1$ with respect to the tangle $T$. 

To finish the definition of the path-model representation, we have
to specify the weight of every compatible path.  There is a
beautiful derivation which yields such weights so that what we get
is indeed a representation (see \cite{ref:Aha05,ref:Aha07} for a
combinatorial exposition of this derivation); here we will not
provide the details but only the resulting definition of the matrix
representation.   We define: $\theta\EqDef \pi/k$, and then
\begin{equation}
\label{def:lambda}
    \lambda_j \EqDef \sin(\pi j/k) = \sin(j\theta) \ ,
\end{equation}
and we have by Equations (\ref{def:d0}),(\ref{def:A0})
\begin{eqnarray}\label{def:d}
  A &= ie^{-i\theta/2} \ .\\
  d &= 2\cos\theta \ ,        \label{def:A}
\end{eqnarray}
We can now define
the matrices $\Phi_i = \rho(E_i)$, and through them $\rho_i =
\rho(\sigma_i)$ by \eref{def:TL-sigma}:
\begin{equation}
\label{def:rho}
  \rho(\sigma_i) = \rho(AE_i + A^{-1}\Id) = A\Phi_i + A^{-1}\Id \ .
\end{equation}
We consider an arbitrary path $p=z_1\to z_2 \to z_3 \to \ldots$,
where $z_i$ is the position on the path \emph{before} taking the
$i$'th step. For brevity, we will denote by $\downdown$ the path in
which the $i$ and $i+1$ steps are descending (i.e., to $z_i-1$ and
then to $z_i-2$), and similarly $\upup$ for two ascending steps.
Similarly, the paths $\updown$, $\downup$ denote a combination of
ascending and descending, and it is agreed that they are coincide
with each other at \emph{all} but the $i,i+1$ steps. Then the
$\Phi_i$ matrices are given by
\begin{eqnarray}
\label{eq:phi-first}
  \Phi_i\ket{\downdown} &=& 0 \ , \\
  \label{eq:phi-second}
  \Phi_i\ket{\downup} &=&
    \frac{\lambda_{z_i-1}}{\lambda_{z_i}}
    \ket{\downup}
    + \frac{\sqrt{\lambda_{z_i+1}\lambda_{z_i-1}}}{\lambda_{z_i}}
       \ket{\updown} \ , \\
  \label{eq:phi-third}
  \Phi_i\ket{\updown} &=&
    \frac{\lambda_{z_i+1}}{\lambda_{z_i}}
    \ket{\updown}
    + \frac{\sqrt{\lambda_{z_i+1}\lambda_{z_i-1}}}{\lambda_{z_i}}
       \ket{\downup} \ , \\
  \Phi_i \ket{\upup} &=& 0 \ .
\label{eq:phi-last}
\end{eqnarray}

Notice that by \eref{def:rho}, the operators $\rho_i$ have the same
invariant subspaces as the $\Phi_i$ operators. Specifically, in the
paths basis, $\rho_i$ breaks into one-dimensional and
two-dimensional blocks (but notice that these are different blocks
for different operators) that consist of 
$\big\{\ket{\downdown}\big\},\big\{ \ket{\upup}\big\}$ paths and
$\big\{\ket{\updown}, \ket{\downup}\big\}$ paths respectively.
We also see that $\Phi_i$, and hence $\rho_i$, does not
change the end point of a path, because they only mix paths that
coincide at all but the $i+1$ site. Therefore, the path
representation breaks into representations over subspaces that
correspond to paths that end at a particular $\ell$. We denote these
subspaces by $H_{n,k,\ell}$, and note that $H_{n,k} =
\sum_{\ell=1}^{k-1} \oplus H_{n,k,\ell}$.

\subsection{From braids to links to Jones Polynomial}

It turns out that there is a very strong connection between the
path-model representation of a braid and the Jones polynomial of its
plat closure. We will not define here the Jones polynomial, but only
refer to it by notation, $V_L(\cdot)$.  The Jones Polynomial of the
plat closure of every $b\in B_n$ can be given by a ``sandwich''
product of the operator $\rho(b)$ with a special vector
$\ket{\alpha}\in H_{n,k}$. Specifically, let $b^{pl}$ denote the
plat closure of the braid $b$, and $V_{b^{pl}}(\cdot)$ its Jones
polynomial. Let $\alpha = 1\to 2\to 1\to 2\to \cdots = \zigzag $
denote the ``zig-zag'' path, and $\ket{\alpha}$ its corresponding
vector. Then the following equality holds:
\begin{equation}
  \label{eq:jones}
  \bra{\alpha}\rho(b)\ket{\alpha} = \frac{1}{\Delta}
    V_{b^{pl}}(A^{-4}) \ ,
\end{equation}
where $\Delta$ is given by 
  \begin{equation}
  \label{def:app-scale}
    \Delta \EqDef d^{n/2-1}(-A)^{3w(b^{pl})} \ .
  \end{equation}  
Here, $A=ie^{-i\pi/(2k)}$ and $d=2\cos(\pi/k)$ are given in
(\ref{def:d0}, \ref{def:A0}), and $w(b^{pl})$ is the \emph{writhe}
of the link $b^{pl}$, which is a trivial function of a link -- it is
basically a sum over all its crossings. $V_{b^{pl}}(A^{-4})$ is the
Jones polynomial of $b^{pl}$, evaluated at $A^{-4} = \exp(-2\pi
i/k)$.  

We note that both the writhe and the Jones polynomial are only
defined for \emph{oriented} links, and therefore we must choose some
orientation for $b^{pl}$ to make the above well defined; it does not
matter, however, which orientation we pick since the combination
$(-A)^{-3w(b^{pl})} V_{b^{pl}}(A^{-4})$ is independent of the
orientation (in agreement with the LHS of \eref{eq:jones}). In fact,
this combination is precisely the \emph{Kauffman bracket} $\langle
b^{pl}\rangle$ \cite{ref:Kau87}, which is also a polynomial of the
link, but we will not use this terminology here.  We further note
that $|\Delta| =d^{n/2-1} = \big(2\cos(\pi/k)\big)^{n/2-1}$. As we
shall see in the following section and in \Sec{sec:poly-k}, this
constant is the approximation scale of our additive approximation.

\section{$\BQP$-hardness for constant $k$}
\label{sec:const-k}

Equation \eref{eq:jones} from the previous section establishes the
connection between the Jones polynomial and a
quantum-mechanical-like expectation value
$\bra{\alpha}\rho(b)\ket{\alpha}$. It is this connection that
enables, on one hand, the approximation of the Jones polynomial by a
quantum computer, and, on the other hand, the simulation of a
quantum computer by approximating the Jones polynomial.

In this section we show the latter result. Specifically, we show
that approximating the Jones polynomial at the $k$'th root of unity
$\exp(2i\pi/k)$ for $k>4, k\ne 6$ is $\BQP$-hard. This result was
already proved by Freedman et al. \cite{ref:Fre02}.  Here and
in the following section, we give our version of the proof, which
uses a somewhat more elementary machinery, and
enables us to prove the $\BQP$-hardness of the $k=\poly(n)$ problem
in \Sec{sec:poly-k}.

For a constant $k$, the exact statement of the result is as follows
\begin{thm}[$\BQP$-hardness for a fixed $k$]
\label{thm:const-k} 
  Let $k>4, k\ne 6$ be an integer, and $t=\exp(2i\pi/k)$ its
  corresponding root of unity. Let $b\in B_n$ be a braid with
  $m=\poly(n)$ crossings, and $b^{pl}$ its plat closure. Finally,
  let $V_{b^{pl}}(t)$ be its Jones polynomial, and $\Delta$ as
  defined in \eref{def:app-scale}
  so that $|\Delta| =
  \big(2\cos(\pi/k)\big)^{n/2-1}$.
  Then given
  a promise that either $|V_{b^{pl}}(t)| \le
  \frac{1}{10}|\Delta|$ or $|V_{b^{pl}}(t)| \ge \frac{9}{10}|\Delta|$, it
  is $\BQP$-hard to decide between the two. 
\end{thm}

The rest of this section is devoted to the proof of this theorem.
The outline of the proof was given in the introduction, and we
repeat it here for readability, and also in order to add a few
missing details.  Fix a $k$, as in Theorem \ref{thm:const-k}.  We
assume we have access to a machine that for given a braid provides
approximations of the Jones polynomial within the same accuracy as
in Theorem~\ref{thm:const-k}, in polynomial time. By \eref{eq:jones}
and the definition of the approximation window $\Delta$ in
\eref{def:app-scale}, this means that we have access to a machine
that given a braid $b$, can decide whether 
$|\bra{\alpha}\rho(b)\ket{\alpha}|$ is larger than $0.9$ or smaller
than $0.1$.  It therefore suffices to reduce a known $\BQP$-hard
problem to this latter approximation problem of
$|\bra{\alpha}\rho(b)\ket{\alpha}|$.

We will do this using the following problem, which is easily shown
to be $\BQP$-hard by standard arguments: Given is a quantum circuit
by its $L$ gates, $U=U_L\cdots U_2\cdot U_1$ on $n$ qubits with
$L=\poly(n)$, decide whether $|\bra{0^{\otimes n}} U \ket{0^{\otimes
n}}| \le \frac{1}{3}$ or $|\bra{0^{\otimes n}} U \ket{0^{\otimes
n}}| \ge \frac{2}{3}$.  This problem is easily seen to remain
$\BQP$-hard even if we assume that the qubits the circuit acts on
are set on a line, and each gate $U_i$ is two-local, acting on
adjacent qubits. 

We will show how given such a quantum circuit, one can efficiently
find a braid $b$ of polynomial number of strands and crossings such
that $\bra{\alpha}\rho(b)\ket{\alpha}$ approximates $\bra{0^{\otimes
n}} U \ket{0^{\otimes n}}$ (say, up to an additive error of $1/10$).
This will suffice to prove Theorem (\ref{thm:const-k}). 

We begin by introducing the Kitaev-Wocjan-Yard 4-steps encoding that
maps strings of bits to paths, and would enable us to map any
quantum gate $U_j$ to an operator on the space of paths.

\subsection{The 4-steps encoding}

In the 4-steps encoding, we encode every bit by a 4-steps path that
starts and ends at the first site:
\begin{eqnarray}
  \ket{\underline{0}} &\EqDef& \ket{1\to 2\to 1\to 2\to 1} 
   = \ket{\ezero} \\
  \ket{\underline{1}} &\EqDef& \ket{1\to 2\to 3\to 2\to 1} =
    \ket{\eone} \ .
\end{eqnarray}
Then a string of $n$ encoded qubits $\ket{x}$ is encoded as a
$4n$-steps path in $H_{4n,k}$, and is denoted by
$\ket{\underline{x}}$. These paths are not arbitrary paths in
$H_{4n, k}$, as they return to the first site every 4 steps. We
denote by $S$ the subspace that is spanned by all these paths. We
note that the zig-zag path $\ket{\alpha}\in H_{4n,k}$ is actually the
encoded string $\ket{\underline{0}}^{\otimes n}$.

Next, just as we encode bit strings, we encode the computational
gates: every gate $U$ is encoded by
\begin{eqnarray}
\label{eq:U}
  \UU = \sum_{i,j} U_{ij} \ket{\underline{i}}\bra{\underline{j}} 
       + \Id_{\mbox{\tiny over rest of space}} \ ,
\end{eqnarray}
where $i,j$ denote bit strings and $\ket{\underline{i}},
\ket{\underline{i}}$ their encoding.  Then the product
$U=U_L\cdot\ldots\cdot U_1$ naturally translates to $\UU =
\UU_L\cdot\ldots\cdot \UU_1$ and so by finding braids $b_i\in
B_{4n}$ such that $\rho(b_i) \simeq \UU_i$ and then taking their
product $b = b_L\cdot\ldots \cdot b_1$, we will get $\rho(b) \simeq
\UU$.  Consequently,
\begin{equation}
\label{eq:app0}
  \bra{0^{\otimes n}}U\ket{0^{\otimes n}} 
    = \bra{\underline{0}^{\otimes n}}\UU\ket{\underline{0}^{\otimes n}}
    \simeq \bra{\underline{0}^{\otimes n}}
      \rho(b)\ket{\underline{0}^{\otimes n}}=
\bra{\alpha}\rho(b)\ket{\alpha}  \ .
\end{equation}
In fact, we will not be so ambitious; we will only require that
$\rho(b_i) \simeq \UU_i$ on the subspace $S$, and show that this
suffices. 
 
The advantage of using this particular encoding is that, together
with the tensorial structure of the qubits, it allows us to
concentrate on the ``reduced'' braid group $B_8$ instead of the
larger group $B_{4n}$. Let us explain exactly what is meant by that.
Suppose that we wish to perform an operation on the $s, s+1$ encoded
qubits of some path $\ket{p}\in S$.  Then we must use a braid $b\in
B_{4n}$ that mixes the 8 strands $4(s-1)+1 \to 4(s+1)$ while being
trivial on the rest. However, since $\ket{p}\in S$, its path reaches
the first site before the $4(s-1)+1$ and $4(s+1)+1$ steps. Therefore
the three partial paths that are defined by the steps $1 \to
4(s-1)$,\quad $4(s-1)+1 \to 4(s+1)$ and $4(s+1)+1\to 4n$ are all
\emph{legitimate} paths over the graph $G_k$ (i.e., they start and
end at the first site and never leave $G_k$). We denote these
partial paths by $p_0, \tilde{p}$ and $p_1$ respectively, and write
$\ket{p}=\ket{p_0}\otimes \ket{\tilde{p}}\otimes\ket{p_1}$. Notice
also that $\ket{\tilde{p}}\in H_{8,k,1}$. We will add a tilde to all
vectors and operators that act on the $H_{8,k,1}$ space. In
particular, we define $\tilde{b}\in B_8$ to be the ``reduced''
version of $b$, created by the $8$ non-trivial strands of $b$. 

It is now easy to verify that
\begin{equation}
\label{eq:reduction}
  \rho(b)\ket{p} = \rho(b)
     \Big(\ket{p_0}\otimes\ket{\tilde{p}}\otimes\ket{p_1}\Big)
     = \ket{p_0}\otimes\Big(\rho(\tilde{b})\ket{\tilde{p}}\Big)
       \otimes\ket{p_1} \ .
\end{equation}
This follows from the definition of the generators $\rho_i$ in
(\ref{def:rho}, \ref{eq:phi-first}-\ref{eq:phi-last}) which only
depend on $z_i$ - the position of the path after $i-1$ steps, and
not on the index $i$ itself. 

By linearity, we can extend \eref{eq:reduction} to all vectors in $S$,
which are simply superpositions of encoded paths. Therefore, as long
as $\ket{p}\in S$, it is enough to search for an appropriate braid
in the much simpler group, $B_8$, instead of looking in the full
$B_{4n}$ group. What remains to show is that (i) we can approximate
any operator on $H_{8,k,1}$ using a $\tilde{b}\in B_8$ (and that
this can be done efficiently), and (ii) that the state we work with
is always sufficiently close to the subspace $S$ where
\eref{eq:reduction} is valid. The next theorem and its subsequent
claim show exactly that.

\begin{thm}[Density and efficiency in $B_8$ for a constant $k$]
  \label{thm:main}   
  Fix $k>4$, $k\ne 6$, and let $\tilde{\UU}$ be an encoded
  two-qubit quantum gate, and $\delta>0$.  Then there exists a braid
  $\tilde{b}\in B_8$, consisting of $\poly(1/\delta)$ generators of
  $B_8$, such that for every $\ket{\tilde{p}}\in H_{8,k,1}$,
  \begin{equation*}
    \norm{(\rho(\tilde{b})- \tilde{\UU})\ket{\tilde{p}}} \le \delta
    \ ,
  \end{equation*}
  that can be found in
  $\poly(1/\delta)$ time.
\end{thm} 
The proof of this theorem is given in \Sec{sec:B8-const}.  Let us
now see how, together with \eref{eq:reduction}, it can be used to
construct the appropriate braid $b\in B_{4n}$ in \eref{eq:app0}. 

Let $U=U_L\cdot\ldots\cdot U_1$ be our quantum circuit, with $U_i$
being local two-qubit gates, and let $\epsilon>0$ be an arbitrary
constant. For every $U_i$ we use the theorem to construct a braid
$\tilde{b}_i\in B_8$, with $\delta=\epsilon/L$, and extend it into a
braid $b_i\in B_n$ by adding identity strands around it. Finally,
$b$ is taken to be the product of these $b_i$'s. We have
\begin{claim} 
  \label{cl:app1}
  $\| \UU_L\cdot \UU_{L-1} \cdots \UU_{1}\ket{\alpha}
   -\rho(b_L)\cdot\rho(b_{L-1})\cdots \rho(b_1)\ket{\alpha}\|\le
   \epsilon. $
\end{claim}

\begin{proof}
  The claim is easily proved by induction. Indeed, assume that 
  \begin{equation}
    \| \UU_{i-1}\cdots \UU_{1}\ket{\alpha}
     - \rho(b_{i-1})\cdots \rho(b_1)\ket{\alpha}\| 
     \le \frac{i-1}{L}\epsilon \ ,
  \end{equation}
  and define $\ket{\beta}\EqDef \UU_{i-1}\cdots
  \UU_{1}\ket{\alpha}$. It is easy to verify that \emph{any} encoded
  gate $\UU$ sends the subspace $S$ into itself and therefore
  $\ket{\beta}\in S$. Consequently
  \begin{equation}
    \| \UU_i\ket{\beta} - \rho(b_i)\ket{\beta}\| = 
    \| \tilde{\UU}_i\ket{\tilde{\beta}} 
       - \rho(\tilde{b}_i)\ket{\tilde{\beta}}\|
     \le \frac{1}{L}\epsilon \ ,
  \end{equation}
  where the first equality follows from the reduction in
  \eref{eq:reduction} and the second inequality follows from the way
  in which we constructed $\tilde{b}_i\in B_8$. Then using the
  induction assumption together with the triangle inequality, we get
  \begin{eqnarray}
    &&\| \UU_i\cdots\UU_1\ket{\alpha} 
    - \rho(b_i)\cdots\rho(b_1)\ket{\alpha}\| \\
    &=& \| \UU_i\ket{\beta} - \rho(b_i)\ket{\beta} 
    + \rho(b_i)\UU_{i-1}\cdots\UU_1\ket{\alpha}
    - \rho(b_i)\cdots\rho(b_1)\ket{\alpha}\| \nonumber \\
    &\le& \|\UU_i\ket{\beta} - \rho(b_i)\ket{\beta}\| 
    + \| \UU_{i-1}\cdots\UU_1\ket{\alpha}
    - \rho(b_{i-1})\cdots\rho(b_1)\ket{\alpha}\| \nonumber\\ 
    &\le& \frac{\epsilon}{L} + (i-1)\frac{\epsilon}{L} 
     = i\frac{\epsilon}{L} \ . \nonumber
  \end{eqnarray}
\end{proof}

This shows that the braid $b=b_L\cdots b_1$ satisfies
\begin{equation}
  \big| \bra{\alpha}\rho(b)\ket{\alpha} 
    - \bra{0^{\otimes n}} U \ket{0^{\otimes n}}\big| \le \epsilon \ .
\end{equation}
Taking $\epsilon=1/10$, and using \eref{eq:jones} then enables us to
decide whether $|\bra{0^{\otimes n}} U \ket{0^{\otimes n}}| \le 1/3$
or $|\bra{0^{\otimes n}} U \ket{0^{\otimes n}}|\ge 2/3$ by deciding
whether $|V_{b^{pl}}(t)| \le \frac{1}{3}\Delta$ or $|V_{b^{pl}}(t)|
\ge \frac{2}{3} \Delta$, as required. Moreover, this procedure is
efficient since by theorem~\ref{thm:main}, the number of braid
generators that are needed to approximate every gate is of the order
$\poly(1/\delta)$, and they can be found in time $\poly(1/\delta)$.
Therefore, overall, as $L=\poly(n)$ and $\delta=\epsilon/L$, $b$ is
made of $\poly(n,1/\epsilon)$ gates and can be found in time
$T=\poly(n,1/\epsilon)$. This concludes the proof of
theorem~\ref{thm:const-k}.

In the next section we will prove theorem~\ref{thm:main} - the $B_8$
density and efficiency theorem for constant $k$.

\section{Proving the $B_8$ density and efficiency theorem}
\label{sec:B8-const}

Our strategy to proving theorem~\ref{thm:main} is to use the famous
Solovay-Kitaev theorem \cite{ref:Kit02a}, which shows that density
implies efficiency. Specifically, we will first prove that the
operators $\rho_1, \ldots, \rho_7$ can approximate any unitary
operator on $H_{8,k,1}$. In other words, we will show that they
generate a dense subgroup in $SU(H_{8,k,1})$. After such density is
proved, the Solovay-Kitaev theorem tells us that it possible to find
a $\delta$-approximation of any unitary $U$ that consists of no more
than $\poly(\log(\delta^{-1}))$ generators in
$\poly(\log(\delta^{-1}))$ steps, thereby proving
theorem~\ref{thm:main}.

The rest of this section is therefore devoted to proving that
$\rho_1, \ldots, \rho_7$ generate a dense subset of $SU(H_{8,k,1})$.
We begin by analyzing the structure of the subspace $H_{8,k,1}$ and
the generators $\rho_1, \ldots, \rho_7$ of the $B_8$ path
representation that act on it.

\subsection{The structure of the generators in $H_{8,k,1}$}
\label{sec:analyze}

\begin{figure}
  \center
   \includegraphics[scale=0.4]{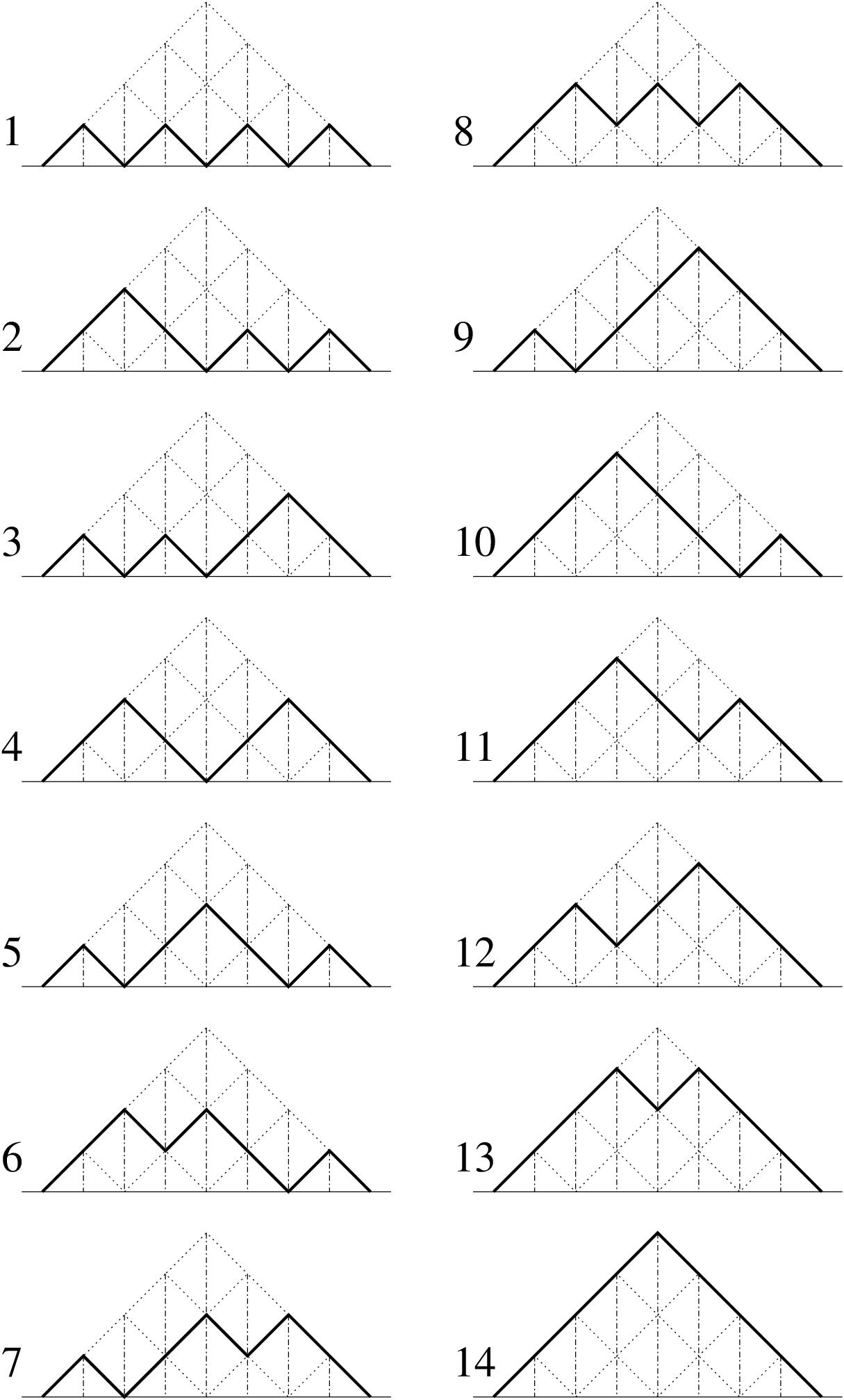}
     \caption{The 14 different vectors that correspond to paths on
    8-strands, starting at 1 and ending at 1}
 \label{fig:8p}
\end{figure}

We begin by noting that for $k>5$, $H_{8,k,1}$ consists of exactly
14 paths\footnote{For $k=5$ there are actually only 13 paths, as
path 14 is illegal (it gets out of the graph). Nevertheless, it is
easy to see that the density proof still holds in that border case.
For $k=4$ we cannot prove density (see theorem~\ref{thm:seed}) while
for $k<4$ the $H_{8,k,1}$ is too small to encode 2 qubits.}, and
hence it is a 14 dimensional space. These paths are labeled by the
numbers $1, \ldots, 14$ and shown graphically in \Fig{fig:8p}.

Let us now describe the structure of the generators on this space.
In \Sec{sec:path-rep} we saw that the generators break into
2-dimensional and 1-dimensional blocks when represented in the
standard basis. Let us look at these blocks in some more detail.

First, by (\ref{eq:phi-first},\ref{eq:phi-last}), $\Phi_i$ nullifies
paths of the form $\ket{\upup}$ and $\ket{\downdown}$, and as a
result they become eigenvectors of $\rho_i$ with an eigenvalue $A^{-1}$.

The $2\times 2$ blocks of $\Phi_i$ mix $\ket{\updown}$ and
$\ket{\downup}$. By (\ref{eq:phi-second}, \ref{eq:phi-third}), these
blocks are
\begin{equation}
  [\Phi_i]_{2\times 2} =
  \twomat{\frac{\lambda_{z_i+1}}{\lambda_{z_i}}}
    {\frac{\sqrt{\lambda_{z_i+1}\lambda_{z_i-1}}}{\lambda_{z_i}}}
    {\frac{\sqrt{\lambda_{z_i+1}\lambda_{z_i-1}}}{\lambda_{z_i}}}
    {\frac{\lambda_{z_i-1}}{\lambda_{z_i}}} \ .
\end{equation}
This matrix has two eigenvalues: $0$ and $2\cos\theta$, and
consequently (by \eref{def:rho}) the eigenvalues of $\rho_i$ in
these blocks are $\{ A^{-1}, -A^{-1}e^{-2i\theta}\}$ -- independent
of $z_i$. In fact, it is not hard to see that all the $\rho_i$
operators are equivalent, namely 
equal under a unitary change of basis. We further notice that
when $z_i=1$, the off-diagonal terms vanish (because
$\lambda_{z_i-1}=\lambda_0=0$), and the blocks become diagonal.

The $2\times 2$ matrix that diagonalizes $[\Phi_i]_{2\times 2}$ (and
consequently $[\rho_i]_{2\times 2}$) is
\begin{equation}
\label{eq:V}
  M(z_i) \EqDef \frac{1}{\sqrt{\lambda_{z_i+1}+\lambda_{z_i-1}}}
  \twomat{\sqrt{\lambda_{z_i+1}}}{-\sqrt{\lambda_{z_i-1}}}
    {\sqrt{\lambda_{z_i-1}}}{\sqrt{\lambda_{z_i+1}}} \ .
\end{equation}
Inside that subspace we have
\begin{equation}
\label{eq:diag-rho}
  [\rho_i]_{2\times 2}
    = A^{-1}\cdot M(z_i)\twomat{-e^{-2i\theta}}{0}{0}{1}M^\dagger(z_i) \ .
\end{equation}

Using the labeling of \Fig{fig:8p}, we write down the block
structure of the seven generators $\rho_1, \ldots, \rho_7$ in
Table~\ref{tab:blocks}. For each operator, the table lists the
non-trivial blocks where $\Phi_i$ does not vanish.  The
one-dimensional blocks correspond to the $z_i=1$ case, and the
two-dimensional blocks correspond to the $z_i>1$ case. 
\begin{table}[htbp]
  \center
    \begin{tabular}[htbp]{llllll} 
        $\rho_1:$ & $(1)$   & $(3)$   & $(5)$    & $(7)$    & $(9)$ \\
        $\rho_2:$ & $(1,2)$ & $(3,4)$ & $(5,6)$  & $(7,8)$  & $(9,12)$ \\
        $\rho_3:$ & $(1)$   & $(3)$   & $(6,10)$ & $(8,11)$ & $(12,13)$ \\
        $\rho_4:$ & $(1,5)$ & $(2,6)$ & $(3,7)$  & $(4,8)$  & $(13,14)$ \\
        $\rho_5:$ & $(1)$   & $(2)$   & $(7,9)$  & $(8,12)$ & $(11,13)$ \\
        $\rho_6:$ & $(1,3)$ & $(2,4)$ & $(5,7)$  & $(6,8)$  & $(10,11)$ \\
        $\rho_7:$ & $(1)$   & $(2)$   & $(5)$    & $(6)$    & $(10)$ 
    \end{tabular}
\caption{The block structure of the generators of $B_8$ in
  $H_{8,k,1}$ for $k>5$.}
\label{tab:blocks}
\end{table}

\subsection{Proving the density}
\label{sec:density}

We will now prove the density part of Theorem~\ref{thm:main}. We
will show that the seven operators $\rho_i$ can approximate any
special unitary matrix on $H_{8,k,1}$, provided that $k>4$ and $k\ne
6$. As it is a 14-dimensional space, we are interested in matrices
$U\in SU(14)$\footnote{For $k=5$ we look at $SU(13)$ and ignore the
vector $14$}.  

We begin by considering the action of $\rho_1$ and $\rho_2$ on this
subspace. From Table~\ref{tab:blocks} we see that these operators
act non-trivially on the five $2\times 2$ blocks $\{1,2\}, \{3,4\},
\{5,6\}, \{7,8\}, \{9,12\}$, while applying the trivial $A^{-1}$
phase on the rest. In these blocks, the $\rho_1$ operator is
represented by $(i)$, whereas the $\rho_2$ operator is represented
by $(i,j)$. Additionally, the operators on all five
blocks are equivalent, namely equal under a unitary 
change of basis. 
The following theorem assures us that in each
such block we may approximate any $SU(2)$ matrix.

\begin{thm}[Jones \cite{ref:Jon83}]\label{thm:seed}
  If $k>4$, and $k\ne 6$, then in each $2\times 2$
  block, the group that is generated by $\rho_1$ and $\rho_2$ is
  dense in $SU(2)$.
\end{thm}

\begin{proof}
  Since $\rho_1$ and $\rho_2$ are not in $SU(2)$, we will look at
  their images under the canonical homomorphism $U(2)\to SU(2)$
  which takes $W\in U(2)$ to $(\det W)^{-1/2}W$, and prove that
  these images form a dense set in $SU(2)$. Then using the fact that
  $[SU(2), SU(2)]=SU(2)$ it will follow that also $\rho_1$ and
  $\rho_2$ generate a dense set in $SU(2)$.

  Let $G=\langle \rho_1, \rho_2 \rangle$ be the group that is
  generated by $\rho_1, \rho_2$. We first use the fact that $G$ is
  infinite as long as $k>2$ and $k\ne 4, 6$. This fact was
  proved by Jones in 1983 and appears in Theorem 5.1
  page 262 in ref \cite{ref:Jon83}. The proof uses the canonical
  homomorphism between $SU(2)$ and $SO(3)$ and the well-known
  classification of all the finite subgroups of $SO(3)$.

  To approximate any element in $SU(2)$ to within an $\epsilon$, we
  pick two matrices in $g_1, g_2\in G$ such that
  $||g_1-g_2||<\epsilon/3$ (we can do that since $G$ has an infinite
  number of elements and $SU(2)$ is compact), and set $g
  \EqDef g_1g_2^{-1}$. Then $\norm{g-\Id}<\epsilon/3$, and
  consequently, if $e^{\pm i\lambda}$ are the eigenvalues of $g$
  then $|e^{\pm i\lambda} - 1| < \epsilon/3$. In addition, $g$ must be
  non-commuting with at least one of the matrices $\rho_1$ or
  $\rho_2$ which we shall denote by $T$. 
  
  Let $U$ be the diagonalizing matrix of $g$:
  $g=U^{-1}\mbox{diag}\{e^{i\lambda}, e^{-i\lambda}\}U$, and define
  the two continuous families of matrices
  \begin{eqnarray}
    R(\phi) &\EqDef& U^{-1}
      \mbox{diag}\{e^{i\phi}, e^{-i\phi}\}U \ , \\
    S(\phi) &\EqDef& \sigma^{-1}R(\phi)\sigma \ .
  \end{eqnarray}
  Then it is easy to see that any matrix $V\in SU(2)$ can be
  presented as the product $R(\alpha) S(\beta) R(\gamma)$ for a
  suitable choice of $\alpha, \beta, \gamma \in \mathbb{R}$ (see,
  for example, Kitaev \cite{ref:Kit02a}). But since
  $|e^{i\lambda}-1|<\epsilon/3$ then any member in the families
  $R(\cdot), S(\cdot)$ can be approximated by multiplications of $g$
  and $\sigma$ up to a distance of $\epsilon/3$, and therefore the
  multiple $R(\alpha) S(\beta) R(\gamma)$ can be approximated to
  within $\epsilon$.
\end{proof}

Next, consider what happens when we are also allowed to act with
$\rho_3$. Looking at Table~\ref{tab:blocks} we see that the
resulting operators are block-diagonal with respect to the blocks
$\{1,2\}$, $\{3,4\}$, $\{5,6,10\}$, $\{7,8,11\}$, $\{9, 12, 13\}$.
Obviously, we can still approximate any $SU(2)$ matrix in the
$2\times 2$ blocks. The next lemma provides a way to increase the
dimensionality of the space on which we have density, in the
following way: suppose we have a set of operators that is dense on
$SU(A)$ and on $SU(B)$, for two orthogonal subspaces $A,B$.
Suppose, in addition, that we have a unitary operator $W$ on
$A\oplus B$ that mixes these two spaces. Specifically, we demand
that there exists a vector $\ket{u}\in A$ such that $W\ket{u}$ has
some non-zero projection on $B$. We call such transformation a
\emph{bridge between A and B}. Then using this bridge, together with
the density on $A$ and $B$, we have density in $SU(A\oplus B)$.  

\begin{lem}[The Bridge Lemma] 
\label{thm:mix} 
  Consider a linear space $C$ which is a direct sum of two
  orthogonal subspaces $A$ and $B$, and assume that $\dim B > \dim A
  \ge 1$. Let $W$ be a bridge transformation between $A$ and $B$ in
  the sense that was defined above.  Then any $U\in SU(C)$ can be
  approximated to an arbitrary precision using a finite sequence of
  transformations from $SU(A)$, $SU(B)$ and $W$. Consequently, the
  group generated by $SU(A)$, $SU(B)$ and $W$ is dense in $SU(C)$.
\end{lem}
\textit{Proof:} Given in Section \ref{sec:tools}.

The bridge lemma implies that it is also possible to approximate any
$SU(3)$ matrix in the $3\times 3$ blocks. As an example, consider
the $\{5,6,10\}$ block. From Theorem~\ref{thm:seed} we already know
that we are able to approximate any $SU(2)$ transformation on the
$\{5,6\}$ block, and by definition, we also have density on the
block $\{10\}$ because it is one dimensional. We may therefore take
the transformation $\rho_3$ as a bridge between these two subspaces
since, for example, it takes the path $10$ into a superposition of
$10$ and $6$. Lemma~\ref{thm:mix} therefore guarantees that together
they can approximate every transformation in $SU(3)$. 

In the above reasoning there are two small cavities that are worth
mentioning, since they will appear in the rest of the proof.
Firstly, the mixing transformation $\rho_3$ is in $U(3)$ rather than
in $SU(3)$.  This, however, is not a real problem, as we can always
consider the transformation $\tilde{\rho}_3 \EqDef c\rho_3$ with $c$
some phase that fixes $\tilde{\rho}_3$ in $SU(3)$. Then $\langle
\rho_1, \rho_2, \tilde{\rho_3}\rangle$ is dense in $SU(3)$, and
since $[SU(N), SU(N)] = SU(N)$ then also $[\langle \rho_1, \rho_2,
\tilde{\rho_3}\rangle, \langle \rho_1, \rho_2,
\tilde{\rho_3}\rangle]$ is dense in $SU(3)$. But the last group is
equal to $[\langle \rho_1, \rho_2, \rho_3\rangle, \langle \rho_1,
\rho_2, \rho_3\rangle]$ since the group bracket cancels out the
phase $c$ and therefore also $\langle \rho_1, \rho_2, \rho_3\rangle$
is dense in $SU(3)$. 

Secondly, we know we can \emph{approximate} any transformation in
$SU(2)$ while Lemma~\ref{thm:mix} assumes that we can get any
transformation in $SU(2)$ precisely. But since the approximation is
made of a \emph{finite} product of operators, all of which can be
approximated as accurately as desired by $\rho_1, \rho_2, \rho_3$,
it follows that we can also approximate any transformation in
$SU(3)$ to any desired accuracy.

Naturally, the next step is to consider what happens when we are
also allowed to act with $\rho_4$. From Table~\ref{tab:blocks} we
see that resulting transformations will be invariant under the
subspaces $\{1, 2, 5, 6, 10\}$, $\{3, 4, 7, 8, 11\}$, $\{9, 12, 13,
14\}$, that together make up the entire 14-dimensional subspace. We
can use Lemma~\ref{thm:mix} again to learn that we can approximate
any $SU(4)$ transformation in the $\{9, 12, 13, 14\}$ block. But
what about the two other, five-dimensional blocks? There we cannot
use Lemma~\ref{thm:mix} directly. To understand why this is so,
consider, for example, the subspace $\{1, 2, 5, 6, 10\}$. We know
that using $\rho_1, \rho_2, \rho_3$ we can approximate any $SU(2)$
transformation on the $\{1,2\}$ block and any $SU(3)$ transformation
on the $\{5,6,10\}$ block. We also know that that $\rho_4$ bridges 
these two blocks. However, to use Lemma~\ref{thm:mix} we must be
able to approximate the $SU(2)$ transformations independently of the
$SU(3)$ transformation. In other words, we must be able to
approximate an $SU(2)$ transformation on the subspace $\{1,2\}$,
while leaving the subspace $\{5,6,10\}$ invariant and vice verse.
But this is not a prior true since the transformations on $\{1,2\}$
are generated by some sequence of the operators $\rho_1, \rho_2,
\rho_3$, which \emph{simultaneously} generates some transformation
on $\{5,6,10\}$. Luckily, we can use the fact that the
dimensionality of the two subspaces is different in order to prove
that such decoupling is possible:
\begin{lem}[The Decoupling Lemma] 
 \label{thm:decouple} Let $G$ be an infinite discrete group, and let
  $A$, $B$ be two finite Linear spaces with different
  dimensionality. Let $\tau_a$ and $\tau_b$ be two homomorphisms of
  $G$ into $SU(A)$ and $SU(B)$ respectively and assume that
  $\tau_a(G)$ is dense in $SU(A)$ and $\tau_b(G)$ is dense in
  $SU(B)$. Then for any $U\in SU(A)$ there exist a series
  $\{\sigma_n\}$ in $G$ such that
  \begin{eqnarray}
    \tau_a(\sigma_n) &\to& U \, \\
    \tau_b(\sigma_n) &\to& \Id \ ,
  \end{eqnarray}
  and vice verse.
\end{lem}
\textit{Proof:}  Given in \Sec{sec:tools}.

It is therefore clear that we are able to approximate any $SU(5)$
transformation on the $\{1,2,5,6,10\}$ and $\{3, 4, 6, 7, 8, 11\}$
blocks. Using $\rho_5$ we can now mix the $\{3, 4, 7, 8, 11\}$
subspace with the $\{9, 12, 13, 14\}$ subspace, and using the fact
that their dimensionality is different, together with
Lemmas~\ref{thm:mix},~\ref{thm:decouple} - we are guaranteed that we
can approximate any $SU(9)$ transformation on the combined
$9$-dimensional subspace.

Finally, by using $\rho_6$, we mix the five-dimensional block
$\{1,2,5,6,10\}$ with the nine-dimensional block from above -
thereby approximating any transformation in $SU(14)$. This completes
the density proof.

\section{$\BQP$-hardness for $k=\poly(n)$}
\label{sec:poly-k}

In this section we prove the central result of this paper. We prove
a stronger version of theorem~\ref{thm:const-k}, in which $k$ is
allowed to depend polynomially on $n$:
\begin{thm}[$\BQP$-hardness for a $k=\poly(n)$]
\label{thm:poly-k} 
  Let $p(\cdot)$ be some polynomial, and let $b\in B_n$ be a braid
  with $m=\poly(n)$ crossings, and $b^{pl}$ its plat closure.
  Finally, let $V_{b^{pl}}(t)$ be its Jones polynomial at
  $t=\exp(2i\pi/k)$ with $k=p(n)$, and define $\Delta$
  as \eref{def:app-scale}, so that $|\Delta| =
  \big(2\cos(\pi/k)\big)^{n/2-1}$.
  Then given the
  promise that either $|V_{b^{pl}}(t)| \le \frac{1}{10}|\Delta|$ or
  $|V_{b^{pl}}(t)| \ge \frac{9}{10}|\Delta|$, it is $\BQP$-hard to
  decide between the two. 
\end{thm}

Looking at the proof of theorem~\ref{thm:const-k}, it is readily
evident that the only obstacle that prevents it to prove also this
case is the fact that we do not know how theorem~\ref{thm:main}
depends on $k$. Specifically, we do not know the dependence of the
running time as well as the length of the resultant braid on $k$.

It is therefore easy to see, that the following stronger version of
theorem~\ref{thm:const-k}, in which both running time and braid
length are polynomial in $k$ would be enough to prove
theorem~\ref{thm:poly-k}:
\begin{thm}[Density and efficiency in $B_8$ for $k=\poly(n)$]
  \label{thm:main-poly-k} 

  Let $k>4$, $k\ne 6$, and let $\tilde{\UU}$ be an encoded
  two-qubit quantum gate, and $\delta>0$.  Then there exists a braid
  $\tilde{b}\in B_8$, consisting of $\poly(1/\delta, k)$ generators of
  $B_8$, such that for every $\ket{\tilde{p}}\in H_{8,k,1}$,
  \begin{equation*}
    \norm{(\rho(\tilde{b})- \tilde{\UU})\ket{\tilde{p}}} \le \delta
    \ ,
  \end{equation*}
  that can be found in
  $\poly(1/\delta, k)$ time.
\end{thm} 
Indeed it is very easy to see that the very same proof of
theorem~\ref{thm:const-k}, but with theorem~\ref{thm:main} replaced
by the above theorem with $k=\poly(n)$, proves
theorem~\ref{thm:poly-k}. The rest of the section, would therefore
be devoted to proving theorem~\ref{thm:main-poly-k}.

{~}

\begin{proof}
  
  As in the proof of theorem~\ref{thm:main}, our main mathematical
  tool is the Solovay-Kitaev theorem. We would like to use the fact
  that for $k>6$, the generators $\rho_1, \ldots, \rho_7$ form a
  dense subset of $SU(14)$, and then use the Solovay-Kitaev
  algorithm to efficiently generate a $\delta$-approximation for any
  given gate.
  
  There is a problem, however, with this simplistic approach. The
  Solovay-Kitaev algorithm contains an initial step, where an
  $\epsilon$-net is constructed; this is a finite set of operators
  that is generated by $\rho_1, \ldots, \rho_7$ and has the property
  that every operator in $SU(14)$ is closer than $\epsilon$ to at
  least one of the elements of the net. $\epsilon$ is a finite
  constant which is unrelated to the target accuracy $\delta$, and
  whose actual value is of the order $10^{-2}$ (see, for example,
  Ref \cite{ref:Chr05}). The existence of such a net is guaranteed
  since we know that $\rho_1, \ldots, \rho_7$ generate a dense set
  in $SU(14)$; its construction time, however, depends on the
  generators. For a fixed set of generators, this is not a problem;
  the construction time becomes a constant. 
  
  However, the situation becomes more tricky when $k$ is no longer
  fixed. The operators $\rho_1, \ldots, \rho_7$ become
  $k$-dependent, and we can no longer treat the $\epsilon$-net
  construction as a constant step. Its complexity must be taken into
  account. The question is therefore whether we can still guarantee
  that the overall computational cost is polynomial in $k$ and in
  $\log(\delta^{-1})$? The answer is positive; this is what will be
  proved in this section. The main observation is that the
  generators $\rho_1, \ldots, \rho_7$ do not behave randomly, but
  rather converge nicely to a $k=\infty$ limit.  The idea of how to
  make use of this fact was explained in the introduction;
  essentially, the idea is that as $k$ becomes larger and larger,
  the generators $\rho_1, \ldots, \rho_7$ do not behave randomly,
  but rather converge nicely to a $k=\infty$ limit. Their dependence
  on $k$ is simple enough so that we can easily approximate $k_0$
  generators, by products of $k$ generators for $k$ which are
  multiplicities of $k_0$.  Therefore we can construct an
  $\epsilon$-net at some large enough yet constant $k_0$, and
  approximate each element in the net by products of high-$k$
  generators, thereby efficiently obtaining an $\epsilon$-net for
  high $k$s. We now provide the details.  

  Let us therefore begin by considering the $k\to \infty$ limit of
  the $\rho_i$ operators. From \Sec{sec:analyze}, we recall that in
  the standard basis these operators decompose into $1\times 1$ or
  $2\times 2$ blocks. The diagonalizing matrix of the $2\times 2$
  blocks is $M_k(z)$, given by \eref{eq:V}, and the eigenvalues are
  $z$-independent, given by $\{A^{-1}, -A^{-1}e^{-2i\theta}\}$ (see
  \eref{eq:diag-rho}).  Here, and in what follows, we explicitly
  added the subscript $k$ to $M(z)$ to indicate its dependence on
  $k$.  

  Notice that up to an overall factor of $A^{-1}$, the eigenvalues
  of the generators are $\{1, -\exp(-2i\pi/k)\}$. So we can express
  the eigenvalues of low $k$'s as products of eigenvalues of high
  $k$'s. However, this is still not enough, as we want the low $k$
  generators themselves to be approximated by products of high $k$
  generators. Luckily, we notice that the diagonalizing matrix
  $M_k(z)$ converges nicely to $M_\infty(z)$ as $k\to \infty$.
  \begin{equation}
    M_\infty(z) \EqDef \lim_{k\to \infty} M_k(z) =
     \frac{1}{\sqrt{2z}}
      \twomat{\sqrt{z+1}}{-\sqrt{z-1}}{\sqrt{z-1}}{\sqrt{z+1}} \ .
  \end{equation}
  We can therefore define an auxiliary low-$k$ generators by taking
  the $k\to \infty$ diagonalizing matrix $M_\infty(z)$ together with
  the eigenvalues at some low $k_0$. Then for high-enough $k$'s, for
  which $M_k(z)$ is close enough to $M_\infty(z)$, we can
  approximate the auxiliary generators by powers of $\rho_i(k)$.

  Specifically, we set $k_0\EqDef 7$, and in accordance with
  \eref{eq:diag-rho}, we define the following set of 7 auxiliary
  generators:
  \begin{equation}
    [\hat{\rho}_i]_{2\times 2}
      = A^{-1}\cdot
        M_\infty(z_i)\twomat{-\exp(-2i\pi/k_0)}{0}{0}{1}M_\infty^\dagger(z_i) \ .
  \end{equation}
  It is easy to see that the auxiliary operators generate a dense
  set in $SU(14)$. Indeed, the density proof in \Sec{sec:density}
  remains valid since it only relies on the eigenvalues of the
  generating operators and on the fact that for $z>1$, $M_k(z)$
  mixes the two standard basis vectors. We will thus generate an
  $\epsilon$-net from $\{\hat{\rho}_i\}$ and use it to generate the
  $\epsilon$-net of $\{\rho_i\}$ for high $k$'s. This is proved in
  the following Lemma:
  \begin{lem}
    Let $\hat{E}$ be an $\epsilon/2$-net, generated from
    $\{\hat{\rho}_i\}$, and assume without loss of generality that
    each element in $\hat{E}$ is a group commutator (this is
    possible since $SU(14)$ is a simple Lie-group and therefore
    $[SU(14), SU(14)]=SU(14)$). Then for large enough $k$, it is
    possible to generate an $\epsilon$-net $E_k$ by replacing every
    occurrence of $\hat{\rho}_i$ in $\hat{E}$ by $(\rho_i)^{2m}$,
    where $\{\rho_i\}$ are the generators at $k$, and $m
    =\mathcal{O}(k)$.
  \end{lem}

  \begin{proof}

    Let $d$ be the maximal number of generators that are needed to
    construct an element in $\hat{E}$. We wish to be able to
    approximate any $\hat{\rho_i}$ up to at least $\epsilon/2d$
    using $\rho_i$. The first thing we take care of is that $M_k(z)$
    will be close enough to $M_\infty(z)$.  We therefore pick an
    integer $K_1$ such that for any $k>K_1$, $\norm{M_k(z) -
    M_\infty(z)} \le \epsilon/(6d)$.
  
    Next, we must find a $K_2$ such that for any $k>K_2$, the
    eigenvalues of $(\rho_i)^{2m}$ will be close enough to
    $\hat{\rho}_i$, for some yet to be determined $m$. This is more
    conveniently done by defining 
    \begin{equation}
      P_i \EqDef A(k_0)\hat{\rho_i} \ , \quad
      Q_i \EqDef [A(k)\rho_i]^2 \ , 
    \end{equation}
    and approximating the operators $P_i$ with $Q_i$. In the end,
    the factors $A(k_0)$, $A(k)$ will cancel out when we plug these
    operators to the group commutator of each element in $\hat{E}$.
    The logic behind these definitions is that these factors cause
    one of the eigenvalues of both $P_i$ and $Q_i$ to be exactly one
    (see \eref{eq:diag-rho}), and therefore we only have to match
    the remaining eigenvalues. Indeed, the non-trivial eigenvalue of
    $P_i$ is $-\exp(-i2\pi/k_0) = \exp(-i\pi(2+k_0)/k_0)$, whereas
    the non-trivial eigenvalue of $Q_i$ is $\exp(-4\pi i/k)$. We
    therefore define
    \begin{equation}
      m\EqDef\left\lfloor\frac{(2+k_0)/k_0}{4/k}\right\rfloor \ ,
    \end{equation}
    and let $K_2$ be such that for every $k>K_2$  
    \begin{eqnarray}
      \label{eq:Delta}  
      \left| e^{-i\pi(2+k_0)/k_0} - e^{-4m\pi i/k}\right| 
       < \epsilon/(6d)\ .
    \end{eqnarray}
    It is easy to see that it is enough to choose $K_2$ (larger than
    $k_0$) for which $|\exp(-4\pi i/K_2) - 1| < \epsilon/(6d)$.
  
    Assume then that $k>\max(K_1, K_2)$ and let us estimate the
    distance between $P_i$ and $(Q_i)^m$. This is the maximal
    distance between the corresponding blocks in the standard basis.
    In the $1\times 1$ blocks both operators have an eigenvalue $1$
    and therefore the distance is zero. In the $2\times 2$ blocks we
    have 
    \begin{eqnarray}
      [P]_{2\times 2} &=& M_\infty^{-1}(z)
       \twomat{e^{-i\pi(2+k_0)/k_0}}{0}{0}{1}M_\infty(z) \ , \\
        {[}Q^m]_{2\times 2} &=& V^{-1}(z)
       \twomat{e^{-4m\pi i/k}}{0}{0}{1}V(z) \ ,
    \end{eqnarray}
    and consequently
    \begin{eqnarray}
      \norm{ [P-Q^m]_{2\times 2}} &\le
        \norm{M^{-1}_\infty(z)-M^{-1}(z)}
          + \left| e^{-i\pi(2+k_0)/k_0} - e^{-4m\pi i/k}\right| \nonumber \\
         &+ \norm{M_\infty(z)-M(z)} \le \epsilon/(2d) \ .
    \end{eqnarray}

    Let us now return to the $\hat{E}$-net and create the $E_k$ net.
    Any element in $\hat{E}$ is a commutator of products of
    $\hat{\rho}_i$, and therefore remains unchanged if we replace
    $\hat{\rho}_i \to P_i$, because the phase factors cancel out in
    the commutator. The distance of this product from a product in
    which we replace $P_i \to (Q_i)^m$ is smaller than $\epsilon/2$
    since $\norm{P_i - (Q_i)^m}<\epsilon/(2d)$ and we have at most
    $d$ terms in the product. The $(Q_i)^m$'s product is unchanged
    upon the replacement $(Q_i)^m \to (\rho_i)^{2m}$ (again, the
    phase factors cancel out), and this results in the $E_k$ net.
    Hence any element in $\hat{E}$ can be efficiently approximated
    up to a distance $\epsilon/2$ by an element of $E_k$. It follows
    that $E_k$ is an $\epsilon$-net.
  \end{proof}

It follows that we can create an $\epsilon$-net from the operators
$\rho_i$, and because $m<k$, the number of steps that are needed to
create this net is bounded by $\poly(k)$.  The next step would be the
application of the Solovay-Kitaev algorithm to approximate any
transformation $U\in SU(14)$ up to an error $\delta$ - and so the
overall computational cost is bounded by $\poly(k, 1/\delta)$ as
required.
  
\end{proof}

\section{General Tools for proving Universality: The Bridge Lemma and 
the decoupling lemma}
\label{sec:tools}

We provide here the proofs of the bridge lemma and the 
decoupling lemma. For convenience, we restate the lemmas. 

\subsection{The Bridge Lemma}
Let us start with redefining a bridge transformation: 

\begin{deff} 
  Given two orthogonal subspaces $A$ and $B$, a unitary operator $W$
  on $A\oplus B$ is said to be a \emph{bridge between A and B} if 
  there exists a vector $\ket{u}\in A$ such that $W\ket{u}$ has some
  non-zero projection on $B$. Note that this notion is symmetric,
  since the existence of such a vector implies the existence of
  $\ket{u}\in B$ such that $W\ket{u}$ has a non-zero projection on
  $A$, by unitarity.  We sometimes say that such a transformation
  \emph{mixes} the two subspaces.  
\end{deff} 

We restate the bridge lemma: 

{~}

\noindent\textbf{Lemma~\ref{thm:mix} (The Bridge Lemma)}
 Consider a linear space $C$ which is a direct sum of two orthogonal
 subspaces $A$ and $B$, and assume that $\dim B > \dim A \ge 1$. Let
 $W$ be a bridge transformation between $A$ and $B$.  Then any $U\in
 SU(C)$ can be approximated to an arbitrary precision using a finite
 sequence of transformations from $SU(A)$, $SU(B)$ and $W$.
 Consequently, the group generated by $SU(A)$, $SU(B)$ and $W$ is
 dense in $SU(C)$.

To prove lemma~\ref{thm:mix} we first need to prove the following
two lemmas:

\begin{lem}
\label{lem:A} Consider a linear space $C$ that is a direct sum of
  two subspaces $A$ and $B$ such that $\dim B > \dim A \ge 1$, and
  let $W\in SU(C)$ be a bridge transformation that mixes the two
  subspaces. Then for every pair of normalized vectors $\ket{\psi},
  \ket{\phi} \in C$, we can approximate a transformation
  $T_{\psi\to\phi}\in SU(C)$ such that $T_{\psi\to\phi}\ket{\psi} =
  \ket{\phi}$, to any desired accuracy using a finite product of
  transformations from $SU(A)$, $SU(B)$ and $W$.
\end{lem}

\begin{proof}
  Instead of approximating the transformation $T_{\psi\to\phi}$ for
  any two vectors $\ket{\psi}, \ket{\phi}$, we will approximate a
  transformation $T_\psi$ that transforms a particular vector
  $\ket{v^*}$ to an arbitrary vector $\ket{\psi}$. Then,
  $T_{\psi\to\phi} = T_\phi T^{-1}_{\psi}$.

  We begin by finding a vector $\ket{v^*}\in B$ for which
  $W\ket{v^*}\in B$. Such vector must exist since $\dim B>\dim A$.
  Indeed, let $\ket{v_1}, \ldots, \ket{v_n}$ be a basis of $B$. Then
  $W\ket{v_i} = \alpha_i\ket{u'_i} + \beta_i\ket{v'_i}$, with
  $\ket{u'_i}\in A$ and $\ket{v'_i}\in B$. Then since $\dim B > \dim
  A$, the $\ket{u'_i}$ vectors are linearly dependent and we can
  find a non-trivial linear combination such that $\sum_i
  c_i\alpha_i\ket{u'_i} = 0$. Then the vector $\ket{v^{*}}\EqDef\sum
  c_i\ket{v_i}$ is in B and $W\ket{v^*}$ has no projection on A.
  
  Next, we pick a vector $\ket{u^*}\in A$ for which $W\ket{u^*}$ has
  some projection on B. Such vector must exist since
  $W$ is a bridge transformation between $A$ and $B$. 
  We write, $W\ket{u^*} = a \ket{u} + b\ket{v}$ with
  $\ket{u}\in A$, $\ket{v}\in B$ and $|a|^2+|b|^2=1$. If $a\neq 0$,
  we find a transformation $U\in SU(A)$ that takes $\ket{u}$ to
  $\ket{u^*}$, and define $\tilde{W} = UW $, otherwise, we set
  $\tilde{W}=W$. We have thus constructed a transformation
  $\tilde{W}$ for which $\tilde{W}\ket{v^*} \in B$ and
  $\tilde{W}\ket{u^*} = a\ket{u^*} + b\ket{v}$ for some $0\le |a| <
  1$.
  
  Now let $\ket{\psi} = \alpha\ket{u_0}+\beta\ket{v_0}$ be an
  arbitrary vector, with $\ket{u_0}\in A$ and $\ket{v_0}\in B$. We
  will now apply a series of unitary operations that will take
  $\ket{\psi}$ closer and closer to $\ket{v^*}$. We start by moving
  $\ket{u_0}$ to $\ket{u^*}$ and $\ket{v_0}$ to $\ket{v^*}$ using
  transformations from $SU(A)$ and $SU(B)$ respectively, yielding
  the vector $\ket{\psi_1} = \alpha\ket{u^*} + \beta\ket{v^*}$.
  Using $\tilde{W}$ we obtain
  \begin{equation}
    \ket{\psi'_1} = \tilde{W}\ket{\psi_1} = \alpha a\ket{u^*} +
    \alpha b\ket{v} + \tilde{W}\ket{v^*} .
  \end{equation}
  As $\ket{v}, \tilde{W}\ket{v^*} \in B$, we can now apply a
  transformation from $SU(B)$ that takes $\alpha b\ket{v} +
  \tilde{W}\ket{v^*}$ to $c_2\ket{v^*}$. Here $c_2$ is the norm of
  $\alpha b\ket{v} + W_2\ket{v^*}$. We obtain
  \begin{equation}
    \ket{\psi_2} = \alpha a\ket{u^*} + c_2\ket{v^*} .
  \end{equation}
  Comparing $\ket{\psi_2}$ to $\ket{\psi_1}=\alpha\ket{u^*} +
  \beta\ket{v^*}$, we see that we managed to move some of the weight
  from $\ket{u^*}$ to $\ket{v^*}$ because $|a|<1$ and both vectors
  are normalized (we only used unitary transformations).  
  
  We now iterate this process: we get $\ket{\psi'_2}$ by applying
  the $W_2$ transformation on $\ket{\psi_2}$, and obtain
  $\ket{\psi_3}$ by moving the $B$ part of $\ket{\psi_2'}$ to
  $\ket{v^*}$. After $n$ such iterations we obtain
  \begin{equation}
    \ket{\psi_n} = \alpha a^{n-1}\ket{u^*} + c_n\ket{v^*} , 
  \end{equation}
  and since $|a|<1$ it is obvious that we exponentially converge to
  $\ket{v^*}$, and in particular we can approximate $T_{\psi}$ (and
  hence $T_{\psi\to\phi}$) to any desired accuracy using a finite
  number of transformations.
\end{proof}

To continue, we need to be able to move a vector from subspace $A$
to subspace B without affecting the rest of the vectors in subspace
$A$. The following Lemma guarantees that this is possible.

\begin{lem}
  \label{lem:B} 
  Under the same conditions of Lemma~\ref{lem:A}, let $\{\ket{u_1},
  \ldots, \ket{u_n}\}$ be an orthonormal basis of $A$ and
  $\{\ket{v_1}, \ldots, \ket{v_m}\}$ be an orthonormal basis of $B$.
  Then using a finite product of transformations from $SU(A), SU(B)$
  and the bridge $W$ between $A$ and $B$, it is possible to
  approximate to any accuracy a transformation $T$ that moves
  $\ket{u_1}$ to $\ket{v_1}$, while leaving the vectors
  $\ket{u_2},\ldots,\ket{u_n}$ unchanged.
\end{lem}

\begin{proof}
For $\dim A = 1$, the problem is trivial since we can simply use
Lemma~\ref{lem:A}. Assume then that $\dim A > 1$, and define the
subspaces $A'\EqDef\mbox{span}\{\ket{u_1} \ldots \ket{u_{n-1}}\}$
and $B'\EqDef\mbox{span}\{\ket{v_2} \ldots \ket{v_m}\}$. By
Lemma~\ref{lem:A} we can approximate a transformation $\tilde{T}$
that takes $\ket{u_n}$ to $\ket{v_1}$. Consider now all the
operators of the form $W = \tilde{T}^{-1}UV'\tilde{T}$ with $U\in
SU(A)$ and $V'\in SU(B')$.  Clearly, $W$ takes $\ket{u_n}$ to itself
- and therefore leaves invariant the subspace $A'\oplus B$. We claim
that there is at least one such transformation, $W^{(1)}$, that also
mixes the subspaces $A'$ and $B$. If this is indeed the case, then
we can repeat the argument for the subspaces $A'$ and $B$, which
together with the particular transformation $W^{(1)}$ satisfy the
conditions of Lemma~\ref{lem:A}. Consequently, we find a
transformation $W^{(2)}$ that takes $\ket{u_{n-1}}$ to $\ket{v_1}$
while leaving $\ket{u_n}$ unchanged and mixing the space that is
spanned by $\ket{u_1},\ldots,\ket{u_{n-2}}$ with $B$. Repeating this
again and again, we are left in the end with a transformation
$W^{(n)}$ that transforms $\ket{u_1}$ to $\ket{v_1}$ and is the
identity over $\ket{u_2}, \ldots, \ket{u_{n}}$. 
Since this recursion has only $n$ steps, it is clear that at any
step we can approximate the mixing transformation $W^{(i)}$ to any
desired accuracy using a finite product of operators from $SU(A)$,
$SU(B)$ and $W$.

Let us now see why $W^{(1)}$ must exist. Indeed, if no such
transformation exists, then for every two operators $U\in SU(A)$ and
$V'\in SU(B')$ there is no mixing between the subspaces $A'$ and
$B$, and therefore there must exist operators $U'\in SU(A')$ and
$V\in SU(B)$ such that
\begin{equation}
  \tilde{T}^{-1}UV'\tilde{T} = U'V \ .
\end{equation}
Then, for every $\ket{u}\in A$ and $\ket{v}\in B$, 
\begin{equation}
  \bra{u}\tilde{T}^{-1}UV'\tilde{T}v\rangle = \bra{u}U'Vv\rangle
    = 0 \ ,
\end{equation}
which implies that for every $U\in SU(A)$ and $V'\in SU(B')$, 
\begin{equation}\label{eq:inter} 
  \bra{u\tilde{T}U}V'\tilde{T}v\rangle = 0 \ .
\end{equation}
Notice that this equation holds also when we take one of the
operators, $U$ or $V'$, to be the identity.  We will use this to
show that $\tilde{T}A=A$ - in contradiction with the fact that
$\tilde{T}\ket{u_1}=\ket{v_1}$.

We first deduce that $\tilde{T}A'\subset A$. To do this we show 
that for all $\ket{u}\in A'$, $\tilde{T}\ket{u}$ has no projection
on $B$. We already know $\tilde{T}\ket{u}$ has zero projection on 
$\ket{v_1}$ (since $\ket{u_1}$ moves to $\ket{v_1}$), so it suffices
to show that $\tilde{T}\ket{u}$ has no projection on $B'$. 

Indeed, by \eref{eq:inter} it would suffice to show that 
$V'\tilde{T}v\rangle$ can be made to be an arbitrary vector in $B'$.
This follows from the following reasoning.  By the same argument as
in the proof of Lemma~\ref{lem:A} there must be a vector
$\ket{v^*}\in B$ such that $\tilde{T}\ket{v^*}\in B$. Moreover,
$\tilde{T}\ket{v^*}$ must be in $B'$ since
\begin{equation}
  \bra{v_1}\tilde{T}v^*\rangle 
   = \bra{v_1\tilde{T}^{-1}}v^*\rangle
   = \bra{u_1}v^*\rangle = 0 \ .
\end{equation}
Since $\tilde{T}\ket{v^*}\in B'$, using an arbitrary $V'\in
SU(B')$, $V'\tilde{T}\ket{v^*}$ can be made to be any vector in $B'$. 

Now pick any $\ket{u^*}\in A'$. Then $\tilde{T}\ket{u^*}\in A$, and
therefore with an arbitrary transformation $U\in SU(A)$,
$U\tilde{T}\ket{u^*}$ can be made to be an arbitrary vector in $A$.
But since for every $\ket{v}\in B$ we have
$\bra{u^*\tilde{T}U}\tilde{T}v\rangle = 0$ then $\tilde{T}\ket{v}$
has no projection on $A$ and consequently, $\tilde{T}B = B$. But
then since $\tilde{T}$ is unitary it follows that $\tilde{T}A=A$ -
which is the contradiction we were seeking. 
\end{proof}

Having proved the last two lemmas, We are now in a position to prove
Lemma~\ref{thm:mix}:

\begin{proof}
Let $\{\ket{u_1}, \ldots, \ket{u_n}\}$ be an orthonormal basis of
$A$ and similarly $\{\ket{v_1}, \ldots, \ket{v_n}\}$ an orthonormal
basis of $B$. We define the following sequence of subspaces 
\begin{equation}
  B_i = B_{i-1}\oplus \ket{u_i} 
\end{equation}
for $i=1 \cdots n$, where $B_0=B$. 

Let us show how to approximate an arbitrary $U\in SU(B_1)$, given
$SU(B)$, $SU(A)$, and $W$.  From Lemma~\ref{lem:B} we can use
$SU(B), SU(A)$ and $W$ to approximate a transformation $T$ that
takes $\ket{u_1}$ to $\ket{v_1}$ while leaving the rest of the
vectors in $A$ intact. Therefore $T\in SU(B_1)$. Now pick an
eigenvector $\ket{\psi}\in B_1$ of $U$ with an eigenvalue
$e^{i\theta}$. Using Lemma~\ref{lem:A} with respect to the subspaces
$span\ket{u_1}, B$ and the mixing transformation $T$, we can
approximate a transformation $W_\psi\in SU(B_1)$ that takes
$\ket{\psi}$ to $\ket{u_1}$. We first show how to approximate the 
transformation $U_1 = W_\psi U W^{-1}_\psi$. 

We notice that $U_1$ has $\ket{u_1}$ as an eigenvector with an
eigenvalue $e^{i\theta}$. Consequently, $U_1$ leaves the subspace
$B$ invariant. Let $V_1$ be the transformation in $SU(B)$ that
satisfies $V_1\ket{v_1}=e^{i\theta}\ket{v_1}$, $V_1\ket{v_2}=
e^{-i\theta}\ket{v_2}$, and leaves the rest of the basis vectors
unchanged. Recalling that $T$ takes $\ket{u_1}$ to $\ket{v_1}$, we
see that $T^{-1}V_1 T$ has $\ket{u_1}$ as an eigenvector with
eigenvalue $e^{i\theta}$ and leaves the subspace $B$ invariant. So
the only difference between $U_1$ and $T^{-1}V_1 T$ is some
transformation $V_2\in SU(B)$, and therefore
\begin{equation}
  U_1 = V_2T^{-1}V_1 T \ ,
\end{equation}
and consequently, 
\begin{equation}
  U = W^{-1}_\psi V_2T^{-1}V_1 T W_\psi \ .
\end{equation}

Now that we have generated all transformations in $SU(B_1)$, we can
generate $SU(B_2)$ using the very same procedure - except now $B_1$
plays the role of $B$ and $\ket{u_2}$ plays the role of $\ket{u_1}$.
In the same method we can work our way all up to $SU(B_n)=SU(C)$. 
\end{proof}

\subsection{The Decoupling Lemma}

{~}

\noindent\textbf{Lemma~\ref{thm:decouple} (The Decoupling Lemma)}
  Let $G$ be an infinite discrete group, and let
  $A$, $B$ be two finite Linear spaces with different
  dimensionality. Let $\tau_a$ and $\tau_b$ be two homomorphisms of
  $G$ into $SU(A)$ and $SU(B)$ respectively and assume that
  $\tau_a(G)$ is dense in $SU(A)$ and $\tau_b(G)$ is dense in
  $SU(B)$. Then for any $U\in SU(A)$ there exist a series
  $\{\sigma_n\}$ in $G$ such that
  \begin{eqnarray}
    \tau_a(\sigma_n) &\to& U \, \\
    \tau_b(\sigma_n) &\to& \Id \ ,
  \end{eqnarray}
  and vice verse.

\begin{proof}
We define two subgroups $H_a \lhd SU(A)$ and $H_b \lhd SU(B)$ by
\begin{eqnarray}
  H_a &\EqDef& \left\{ U\in SU(A) \Big| \  \exists \{\sigma_n\} \quad
    \mbox{s.t.}\quad
         \begin{array}{ccc}
            \tau_a(\sigma_n) &\to& U \\
            \tau_b(\sigma_n) &\to& \Id
         \end{array}
       \right\} \ , \\
  H_b &\EqDef& \left\{ V\in SU(B) \Big| \  \exists \{\sigma_n\} \quad
    \mbox{s.t.}\quad
         \begin{array}{ccc}
            \tau_a(\sigma_n) &\to& \Id \\
            \tau_b(\sigma_n) &\to& V
         \end{array}
       \right\} \ .
\end{eqnarray}
The theorem will be proved once we show that $H_a=SU(A)$ and
$H_b=SU(B)$.

To do that, we first observe that both $H_a$ and $H_b$ are normal
subgroups. It is also straightforward to see that they are closed.
Consider, for example, $H_a$ in $SU(A)$: assume that $\{U_k\}$ in
$H_a$ converges to $U\in SU(A)$. Then there exist series
$\sigma_n^{(k)}$ such that
\begin{eqnarray}
  \lim_{n\to \infty}\tau_a(\sigma_n^{(k)}) &= U_k \ , \\
  \lim_{n\to \infty}\tau_b(\sigma_n^{(k)}) &= \Id \ .
\end{eqnarray}
Without loss of generality, we may choose the series such
that for every $n,k$, 
\begin{equation}
  \left\| \tau_b\left(\sigma^{k}_n\right) - U_k\right\| < 1/n \quad, \quad
  \left\| \tau_b\left(\sigma^{k}_n\right) - \Id\right\| < 1/n \ .
\end{equation}
Then the series 
$\tau_a(\sigma_k^k)\to U$, and we are guaranteed that
$\tau_b(\sigma_k^k)\to\Id$. 

Now, any non-trivial normal subgroup of $SU(N)$ must be
finite\footnote{This follows from the fact that the quotient group
$SU(N)/Z(SU(N))$ is a simple group, and $Z(SU(N))$ - the center of
$SU(N)$ - is finite. See for example Theorem $11.26$, at page 108
of Ref~\cite{ref:Gro01}}. Therefore if we show that $H_a$ and $H_b$
are infinite it will follow that $H_a = SU(A)$ and $H_b=SU(B)$. To
do that, we first show that there is an isomorphism of groups $M$ (a
$1-1$ and onto mapping which preserves the action of the group)
between the coset groups $SU(A)/H_a$ and $SU(B)/H_b$. We define $M$
as follows: $M(UH_a)=VH_b$ if there exists a series $\{\sigma_n\}$
such that
\begin{equation}
  \tau_a(\sigma_n) \to U \ , \quad \tau_b(\sigma_n) \to V \ .
\end{equation}
We first need to show that this function is well defined 
for all cosets $UH_a$. 
This follows from: 
\begin{itemize}
  \item For each coset $UH_a$, there exists at least 
  one series that satisfies the requirements of the definition of $M$.  
  Indeed, pick a series $\{\sigma_n\}$
  such that $\tau_a(\sigma_n)\to U$. Then the series
  $\tau_b(\sigma_n)$ in $SU(B)$ must have a limiting point since
  $SU(B)$ is compact. Therefore there exists a sub-series
  $\{\sigma_{n_k}\}$ such that $\tau_a(\sigma_{n_k})\to U$ and
  $\tau_b(\sigma_{n_k})\to V$. We have $M(UH_a)=VH_b$
  
  \item $M(UH_a)$ is defined uniquely. 
   Indeed, assume there exist two series
  $\{\sigma^{(1)}_n\}$ and $\{\sigma^{(2)}_n\}$ such that
  \begin{eqnarray}
    \tau_a(\sigma^{(1)}_n) \to U_1 \ , \quad
    \tau_b(\sigma^{(1)}_n) \to V_1 \ , \\
    \tau_a(\sigma^{(2)}_n) \to U_2 \ , \quad
    \tau_b(\sigma^{(2)}_n) \to V_2 \ , 
  \end{eqnarray}
  with $U_1$ and $U_2$ in the coset $UH_a$. We will show that $V_1$
  and $V_2$ must be in the same coset of $H_b$.  Denote
  $\Delta\EqDef U_1U_2^{-1}$.  Since $H_a$ is normal, $\Delta$ is in
  $H_a$ and we may therefore find a series $\{\sigma^{(3)}_n\}$ such
  that 
  \begin{equation}
    \tau_a(\sigma^{(3)}_n) \to U_1U_2^{-1} \ , \quad
    \tau_b(\sigma^{(3)}_n) \to \Id \ . 
  \end{equation}
  Then looking at the series $\sigma^{(4)}_n \EqDef
  \big(\sigma^{(1)}_n\big)^{-1}\sigma^{(3)}_n \sigma^{(2)}_n$, we
  find that
  \begin{equation}
    \tau_a(\sigma^{(4)}_n) \to U_1^{-1}U_1U_2^{-1}U_2 
      = \Id \ , \quad
    \tau_b(\sigma^{(4)}_n) \to V_1^{-1}V_2 \ . 
  \end{equation}
  Therefore $V_1^{-1}V_2 \in H_b$, and so $V_1H_b=V_2H_b$. 
\end{itemize}

It remains to show that $M$ is $1-1$, onto, and preserves the action 
of the group. The onto part follows if we start with a series that 
converges to some $V$ in $VH_b$ and apply the same reasoning as in the 
first item above. The $1-1$ part follows if we apply the same reasoning 
as in the second point above, starting with the $V_1,V_2$ in the same coset 
instead of the opposite direction. The fact that $M$ is a homomorphism 
follows from the fact that $\tau$ is a representation.   

Recall now that $H_a$ and $H_b$ can be either finite groups or equal
to their ``supergroup''. So there are four possibilities: 
\begin{enumerate}
  \item $H_a$ is finite and $H_b=SU(B)$.
  \item $H_b$ is finite and $H_a=SU(A)$.
  \item Both $H_a$ and $H_b$ are finite.
  \item $H_a=SU(A)$ and $H_b=SU(B)$.
\end{enumerate}
The first and second cases are impossible since, for example, if
$H_a$ is finite and $H_b=SU(B)$ then $SU(B)/H_b$ has only one coset
while $SU(A)/H_a$ has infinitely many - and thus they cannot be
related by a $1-1$ onto map.

Let us now see why the third case is also impossible. To do this, we
will show that $M$ is a continuous map. Indeed, assume that $U_k H_a
\to U H_a$ (here, convergence means that for some representatives of
the cosets we have $U_k \to U$. It is easy to check that this is
well defined). Then let $M(U_kH_a)=V_kH_b$, $M(UH_a)=VH_b$. We will
show that $V_kH_b \to VH_b$. The proof is straightforward, similarly
to the proof that $H_a$ is closed.  pick pick a series of series
$\{\sigma^{(k)}_n\}$ such that 
\begin{eqnarray}
  \lim_{n\to \infty}\tau_a(\sigma_n^{(k)}) &=& U_k \ , \\
  \lim_{n\to \infty}\tau_b(\sigma_n^{(k)}) &=& V_k \ ,
\end{eqnarray}
and without any loss of generality we assume that 
\begin{equation}
  \left\| \tau_a\left(\sigma^{k}_n\right) - U_k\right\| < 1/n~~~,~~~ 
  \left\| \tau_b\left(\sigma^{k}_n\right) - V_k\right\| < 1/n \ .
\end{equation}
Then since $U_k\to U$, we have $\tau_a(\sigma^{k}_{k}) \to U$, and
since $M(UH_a)=VH_b$ we can find a sub-series
$\tau_b(\sigma^{k_\ell}_{k_\ell})$ that converges to some
$\tilde{V}\in VH_b$. In order not to overload the notation, let us
re-define $k$ to be that sub-series. We now claim $V_k \to
\tilde{V}$. Indeed, for each $\epsilon>0$, we can choose $K$ such
that for each $k>K$, $1/k < \epsilon/2$ and $\|
\tau_b\left(\sigma^k_{k}\right) - \tilde{V}\| < \epsilon/2$. Then 
\begin{equation}
  \| V_k - \tilde{V} \| \le   
  \| V_k - \tau_b\left(\sigma^k_{n_k}\right)\| +
  \| \tau_b\left(\sigma^k_{n_k}\right) - \tilde{V}\|
  \le \epsilon \ .
\end{equation}

Now $H_a$ and $H_b$ are closed normal subgroups, and therefore
$SU(A)/H_a$ and $SU(B)/H_b$ are Lie groups themselves (see for
example, Theorem 3.64, pp 124, in \cite{ref:War83}). Furthermore,
since every continuous homomorphism between Lie groups is also
smooth (see for example, Theorem 3.39, pp 109, in
\cite{ref:War83}), we have found a smooth diffeomorphism ($1-1$
homeomorphism) between two differentiable manifolds. However, since
both $H_a$ and $H_b$ are finite then 
\begin{equation}
  \dim SU(A)/H_a = \dim SU(A) \ne \dim SU(B) = \dim SU(B)/H_b \ ,
\end{equation}
and it is therefore impossible to find a diffeomorphism between the
two manifolds.

\end{proof}


\section{Acknowledgments}
The authors wish to thank an anonymous referee for spotting a couple of  
errors in the parameters of the main theorems. 



\begin{thebibliography}{99}

\bibitem{ref:Aar05} Aaronson, S., ``\emph{Quantum computing,
  postselection, and probabilistic polynomial-time}'',
  Proc.~R.~Soc.~A \textbf{8} vol. 461 no. 2063 pp. 3473--3482,
  (2005), \texttt{arXiv:quant-ph/0412187}


\bibitem{ref:Aha06} Aharonov, D. and Arad, I. ``\emph{The
$\BQP$-hardness of approximating the Jones Polynomial}'', (2006),
\texttt{arXiv:quant-ph/0605181}, 


\bibitem{ref:Aha07} Aharonov, D. and Arad, I., ``\emph{Polynomial
  Quantum Algorithms for Additive approximations of the Potts model
  and other Points of the Tutte Plane}'', (2007),
  \texttt{arXiv:quant-ph/0702008}



\bibitem{ref:Aha97} Aharonov,~D., Ben-Or,~M., ``\emph{Fault-tolerant
  quantum computation with constant error}'', Proceedings of the
  twenty-ninth annual ACM symposium on Theory of computing, pp.
  176--188 (1997)

\bibitem{ref:PromiseBQP} Janzing, D. and Wocjan,P. ``\emph{A simple
  $\PBQP$-complete matrix problem}'', Theory of Computing,
  3, pp.
  61--79 (2007)

\bibitem{ref:PromiseBPP} Goldreich, O., ``\emph{On promise
  problems}'', Electronic Colloquium on Computational Complexity,
  18, (2005).\\
  \texttt{http://eccc.hpi-web.de/eccc-reports/2005/TR05-018/index.html}


\bibitem{ref:FT} Aharonov, ~D. and Ben-Or, ~M. ``\emph{Quantum
  Computation with Constant Error Rate}''. SIAM J. Comput. 38(4):
  pp. 1207--1282, (2008)

\bibitem{ref:Aha05} Aharonov~D., Jones~V.~F, Landau~Z., in proceedings of
  the 38th ACM Symposium on Theory of Computing (STOC 2006) 
  Seattle, Washington, USA. \texttt{arXiv:quant-ph/0511096}


\bibitem{ref:Gor10} Alagic, G. and Jordan, S.~P. and K{\"o}nig, R. 
  and Reichardt, B.~W., ``\emph{Estimating Turaev-Viro
  three-manifold invariants is universal for quantum computation}'', 
  Phys. Rev. A 82, 040302(R) (2010), \texttt{ref:arXiv:1003.0923}


\bibitem{ref:Ara10} Arad~I., Landau~Z., ``\emph{Quantum Computation and
the Evaluation of Tensor Networks}'', SIAM J. Comput. \textbf{39}, 7, 
pp. 3089--3121 (2010), \texttt{arXiv:0805.0040}

\bibitem{ref:Art25} Artin~E., Theorie der Zopfe, Hamburg Abh.~,
  (1925), 4, pp 47-72


\bibitem{ref:Bac00} Bacon, D.  and Kempe, J.  and Lidar, D. A. and
  Whaley, K. B, ``\emph{Universal Fault-Tolerant Quantum Computation
  on Decoherence-Free Subspaces}'', Phys.~Rev.~Lett. \textbf{85},
  pp. 1758--1761 (2000)

\bibitem{ref:Bac01} Bacon, D.  and Kempe, J. and {DiVincenzo}, D.~P.
  and Lidar, D. A. and Whaley, K. B, ``\emph{Encoded Universality in
  Physical Implementations of a Quantum Computer}'', Proc. of
  the 1st International Conference on Experimental Implementations
  of Quantum Computation, Sydney, Australia, edited by R. Clark
  (Rinton, Princeton, NJ, 2001), pp. 257, \texttt{arXiv:quant-ph/0102140}


\bibitem{ref:bv} Bernstein E. and Vazirani U., 
Quantum complexity theory, 
Proceedings of Symposium on the Theory of Computing (STOC 1993), 1993. 
Journal version: 
Special issue on Quantum Computation of the Siam Journal of Computing, 
Oct. 1997. 

\bibitem{ref:Bor05} Bordewich~M., Freedman~M., Lovasz~L., Welsh~D.,
  "\emph{Approximate counting and quantum computation}",
  Combinatorics, Probability and Computing, \textbf{14}, pp
  737--754, (2005).


\bibitem{ref:Chr05} Dawson,~C.~M. and Nielsen, ~M. ~A., 
  ``\emph{The Solovay-Kitaev algorithm}'', 
  \texttt{arXiv:quant-ph/0505030}

\bibitem{ref:Fre98} Freedman~M.~H., {\emph P/NP and the quantum field
  computer}, Proc. Natl. Acad. Sci., USA, \textbf{95}, (1998), 98--101


\bibitem{ref:Fre02b}
  Freedman~M.~H., Kitaev~A., Wang~Z.,
  ``\emph{Simulation of topological field theories by quantum
  computers}'',
  Commun.~Math.~Phys. \textbf{227} pp. 587--603, (2002)

\bibitem{ref:Fre02} Freedman~M.~H., Larsen~M., Wang~Z. ``\emph{A Modular
  Functor which is Universal for Quantum Computation}'',
  Commun.~Math.~Phys., \textbf{227}, pp. 605--622, (2002)

\bibitem{ref:Fre03} Freedman~M.~H., Kitaev~A., Larsen~M., Wang~Z., 
  \emph{Topological quantum computation. Mathematical challenges of
  the 21st century} (Los Angeles, CA, 2000). Bull. Amer. Math. Soc.
  (N.S.) 40, no. 1, pp. 31--38, (2003)



\bibitem{ref:Sil06} Garnerone~S., Marzuoli~A., Rasetti~M.,
  ``\emph{Quantum automata, braid group and link polynomials}'', 
  2006, \texttt{arXiv:quant-ph/0601169}

\bibitem{ref:GJ} Goldberg~L.~A., and Jerrum~M.,  
Inapproximability of the Tutte polynomial, 
Infomation and Computation 206(7), 908-929 (July 2008)


\bibitem{ref:Gro01} Grove,~L. ~C. ``\emph{Classical Groups and
  Geometric Algebra''}, Graduate Studies in Mathematics; V~39, 
  American Mathematical Society, (2001).


\bibitem{ref:Jae90} Jaeger F, Vertigan D. L., Welsh D. J. A., 
``\emph{On the computational complexity of the Jones and
Tutte polynomials}'',  Math.~Proc.~Cambridge~Philos.~Soc. \textbf{108},
no.~1, pp~35--53, (1990).


\bibitem{ref:Jon83} Jones~V.~F.~R, ``Braid groups, Hecke algebras
  and type $II_1$ factors''. In: Geometric methods in operator
  algebras, Proc. of the US-Japan Seminar, Kyoto, July 1983. See
  Theorem 5.1, page 262.

\bibitem{ref:Jon85} Jones~V.~F.~R, ``\emph{A polynomial invariant
  for via Von Neumann algebras}'', Bull. Amer. Math. Soc
  (N.~S.),  \textbf{12}, pp.~103--111, (1985)

\bibitem{ref:Jon86} Jones~V.~F.~R. , ``\emph{Braid groups, Hecke 
  Algebras and type II factors}'', in Geometric methods in 
  Operator Algebras, Pitman Research Notes in Math., \textbf{123} ,
  pp.~242--273, (1986)

\bibitem{ref:Kau87} Kauffman~L., ``State models and the Jones
  polynomial'', Topology, \textbf{26}, pp. 395-407, (1987).

\bibitem{ref:Kem01} Kempe, J. and Bacon, D. and {DiVincenzo}, D.~P.
  and Whaley, K. B, ``\emph{Encoded Universality from a Single
  Physical Interaction}'', QIP, \textbf{1} (Special Issue) 
  pp.~33--55, (2001), \texttt{arXiv:quant-ph/0112013}

\bibitem{ref:Kit05} 
  Kitaev~A.~Yu., private communication, (2005)

\bibitem{ref:Kit02a} Kitaev~A.~Yu., Shen~A.~H., Vyalyi~M.~N~,
  ``Classical and quantum computation'', vol 47 of \emph{Graduate
  Studies in Mathematics}. Amsterdam Mathematical Society, Providence,
  Rhode Island, (2002).

\bibitem{ref:Gre09} Kuperberg, G. ``\emph{How hard is it to
approximate the Jones polynomial?}'', \texttt{arXiv:0908.0512} (2009)

\bibitem{ref:Sho08}  Shor, Peter~W. and Jordan, Stephen~P.,
``\emph{Estimating Jones polynomials is a complete problem for one
clean qubit}'', Quantum Information and Computation, \textbf{8},
pp~681, (2008), \texttt{arXiv:0707.2831v3}



\bibitem{ref:Tem71} Temperley, H.~N.~V. and Lieb, E.~H., 
  ``\emph{Relations between the percolation and colouring problem
  and other graph-theoretical problems associated with regular
  planar lattices: some exact results for the `Precolation'
  problem}''', in Proceedings of the Royal Society of London. Series
  A, \textbf{322}, pp 251--280, (1971)

\bibitem{ref:War83} Warner~F.~W., ``Foundations of Differentiable
  Manifolds and Lie Groups'', Springer, (1983) (Second edition).


\bibitem{ref:Wit89} Witten~E., ``Quantum Field Theory and the Jones
  Polynomial'', Commun. Math. Phys., \textbf{121}, pp.~351-399,
  (1989).


\bibitem{ref:Woc06c} Wocjan~P., Janzing~D., ``\emph{Estimating
diagonal entries of powers of sparse symmetric matrices is
$\BQP$-complete}'', (2006), \texttt{arXiv:quant-ph/0606229} 

\bibitem{ref:Woc07} Wocjan~P., Janzing~D., ``\emph{A
  PromiseBQP-complete String Rewriting Problem}'', (2007), 
  \texttt{arXiv:0705.1180} 



\bibitem{ref:Woc06a} Wocjan~P., Yard~J. ``The Jones polynomial: quantum 
  algorithms and applications in quantum complexity theory'', (2006), 
  \texttt{arXiv:quant-ph/0603069}

\bibitem{ref:Woc06b} Wocjan~P.,  Zhang~S., ``\emph{Several natural
$\BQP$-Complete problems}'', (2006), \texttt{arXiv:quant-ph/0606179} 



\end{thebibliography}
\end{document}